\documentclass[%
 reprint,
superscriptaddress,
 amsmath,amssymb,
 aps,pre,hyphens,floatfix
]{revtex4-2}

\usepackage[]{amsmath}
\usepackage{mathrsfs}
\usepackage{mathtools} 
\usepackage{enumerate}
\usepackage[usenames,dvipsnames,svgnames,table]{xcolor}
\usepackage{IEEEtrantools}
\usepackage{graphicx}
\usepackage{dcolumn}  
\usepackage{bm}       
\usepackage{array}    
\usepackage{multirow}
\usepackage{soul}
\usepackage[normalem]{ulem}
\usepackage[T1]{fontenc}
\usepackage{subfigure}
\usepackage{diagbox}

\newcommand{\stkout}[1]{\ifmmode\text{\sout{\ensuremath{#1}}}\else\sout{#1}\fi}
\newcommand{\mycomment}[1]{}
\newcommand{\stk}[1]{\ifmmode\text{\sout{\ensuremath{#1}}}\else\sout{#1}\fi}
\begin{document}

\title{Supervised and unsupervised learning with numerical computation \\for the Wolfram cellular automata
}

\author{Kui Tuo}
\affiliation{Key Laboratory of Quark and Lepton Physics (MOE) and Institute of Particle Physics,\\ Central China Normal University,
Wuhan 430079, China}

\author{Shengfeng Deng}
\affiliation{School of Physics and Information Technology, Shannxi Normal University, Xi’an 710061, China}
\author{Yuxiang Yang}
\affiliation{Key Laboratory of Quark and Lepton Physics (MOE) and Institute of Particle Physics,\\ Central China Normal University,
Wuhan 430079, China}

\author{Yanyang Wang}
\affiliation{Key Laboratory of Quark and Lepton Physics (MOE) and Institute of Particle Physics,\\ Central China Normal University,
Wuhan 430079, China}

\author{ Qiuping A. Wang}
\affiliation{SCIQ Lab, ESIEA, 74 bis Avenue Maurice Thorez, 
94200 Ivry sur Seine}
\author{Wei Li}
\email[]{liw@mail.ccnu.edu.cn}
\affiliation{Key Laboratory of Quark and Lepton Physics (MOE) and Institute of Particle Physics,\\ Central China Normal University,
Wuhan 430079, China}
\affiliation{SCIQ Lab, ESIEA, 74 bis Avenue Maurice Thorez, 
94200 Ivry sur Seine}

\author{Wenjun Zhang}
\email[]{wenjun@ahtcm.edu.cn}
\affiliation{School of Medical Information Engineering, Anhui University of Chinese Medicine, Hefei 230012, China}

\begin{abstract} 
The local rules of Wolfram cellular automata with one-dimensional three-cell neighborhoods are represented by eight-bit binary that encode deterministic update rules. These automata are widely utilized to investigate self-organization phenomena and the dynamics of complex systems. In this work, we employ numerical simulations and computational methods to investigate the asymptotic density and dynamical evolution mechanisms in Wolfram automata. We apply both supervised and unsupervised learning methods to identify the configurations associated with different Wolfram rules. Furthermore, we explore alternative initial conditions under which certain Wolfram rules generate similar fractal patterns over time, even when starting from a single active site. Our results reveal the relationship between the asymptotic density and the initial density of selected rules. The supervised learning methods effectively identify the configurations of various Wolfram rules, while unsupervised methods like principal component analysis and autoencoders can approximately cluster configurations of different Wolfram rules into distinct groups, yielding results that align well with simulated density outputs.

\end{abstract}

\maketitle

\section{Introduction}

Cellular automata (CA) are dynamical systems discrete in time, space and state. The Game of Life \cite{conway1970game} represents a canonical cellular automaton that investigates how simple rules evolve within discrete spaces and whether such rules can generate dynamics simulating lifelike phenomena. However, its limitation lies in its reliance a single fixed rule set. In contrast, Wolfram cellular automata enable the exploration of emergent complexities across diverse deterministic rules \cite{wolfram1983statistical}, revealing how intricate phenomena arise from elementary local interactions. While Wolfram automata operate under deterministic constraints, the Domany-Kinzel (DK) model generalizes this framework as a stochastic cellular automaton \cite{domany1984equivalence}, facilitating the study of non-equilibrium dynamics and phase transitions in disordered systems. Cellular automata are widely used in various fields, such as fluid mechanics \cite{ilachinski2001cellular},
ecosystem modeling \cite{Grimm2005}, simulation of any computer algorithm \cite{rendell2002turing}, urban planning \cite{batty1997cellular}, simulating traffic flow \cite{nagel1992cellular}, mathematics \cite{yacoubi2008mathematical}, stock market \cite{bartolozzi2004stochastic}, crystal growth models \cite{zhao2009cellular}, biological modeling \cite{ermentrout1993cellular}.

Machine learning approaches have garnered significant interest in recent years and have been extensively adopted across diverse domains, such as
 computer vision \cite{lai2019comparison, pathak2018application}, natural language processing \cite{Hinton}, recommendation systems \cite{da2020recommendation}, finance \cite{hansen2020virtue}, healthcare \cite{catacutan2024machine,whalen2022navigating}, large language models (LLMs) \cite{du2023guiding}.

Machine learning, leveraging its robust capabilities in high-dimensional data processing and nonlinear modeling, has been extensively applied to research in physics. For example, in high-energy physics, particle swarm optimization and genetic algorithms have been employed to autonomously optimize hyperparameters of machine learning classifiers in high energy physics data analyses \cite{tani2021evolutionary},  HMPNet integrates HaarPooling with graph neural networks to boost quark-gluon tagging accuracy \cite{ma2023jet}. In astrophysics, a text-mining-based scientometric analysis has been conducted to map the application trends of machine learning in astronomy  \cite{rodriguez2022application}, a machine learning model has been developed to predict cosmological parameters from galaxy cluster properties \cite{qiu2024cosmology}. In quantum simulation, an ab initio machine learning protocol has been developed for intelligent certification of quantum simulators \cite{xiao2022intelligent}, a quantum machine learning framework has been developed for resource-efficient dynamical simulation with provable generalization guarantees \cite{gibbs2024dynamical}, a noise-aware machine learning framework has been developed for robust quantum entanglement distillation and state discrimination over noisy classical channels \cite{chittoor2023quantum}. 

Machine learning has enabled groundbreaking advances in identifying distinct phases of matter. Carrasquilla and Melko's seminal 2016 study demonstrated that supervised machine learning methods can classify ferromagnetic and paramagnetic phases in the classical Ising model \cite{Carrasquilla}, accurately extracting critical points and spatial correlation exponents. This pioneering work catalyzed widespread adoption of machine learning for analyzing diverse phase transitions across condensed matter systems. Machine learning has also been applied to identify more complex phase transitions, such as in three-dimensional Ising models \cite{Zhang3DIsing}, percolation phase transitions \cite{hu2023universality}, topological phase transitions \cite{kaming2021unsupervised, holanda2020machine}, the nonequilibrium phase transitions in the Domany-Kinzel
automata \cite{tuo2024supervised}, and the nonequilibrium phase transitions even-offspring branching annihilating random walks \cite{wang2024supervised}.

In this work, we study the (1+1)-dimensional Wolfram automata by numerical computation and machine learning. Wolfram automata encompass diverse evolution rules, enabling in-depth analysis of dynamical evolution mechanisms in cellular systems. Simulations and numerical computations are employed to investigate the fractal structures, transient time to reach steady states, and asymptotic density of Wolfram automata. Manual analysis struggles to distinguish configurations under complex Wolfram rules, whereas machine learning leverages its image-classification capabilities to automate Wolfram automata identification. By applying both supervised and unsupervised learning methods, a trained neural network can accurately identify diverse configurations across different Wolfram rules after training on a limited set of automata samples. In addition, two unsupervised learning methods, principal component analysis (PCA) and autoencoder, are utilized to classify configurations and estimate the density of Wolfram automata.

The remainder of this paper is organized as follows: In Sec.~\ref{sec:Wolfram_CA}, we briefly introduce the Wolfram automata. Sec.~\ref{sec:Wolfram numerical} presents the numerical computations of (1+1)-dimensional Wolfram automata, which are employed to investigate the fractal structures and asymptotic density of Wolfram automata. Sec.~\ref{sec:Wolfram Supervised} presents the supervised learning of (1+1)-dimensional Wolfram automata, which are identified distinct configurations of Wolfram rules. Sec.~\ref{sec:Wolfram Unsupervised} is about the unsupervised learning results of (1+1)-dimensional Wolfram automata, via autoencoder and PCA. Sec.~\ref{sec:Summary} summarizes the main findings of this work.

\section{Wolfram cellular automata \label{sec:Wolfram_CA}}

Wolfram cellular automata operate based on evolution rules governed by deterministic dynamics, as noted in prior studies \cite{wolfram1983statistical}. Despite their simplicity, these rules can generate complex phenomena. In one-dimensional Wolfram automata, each site (or cell) assumes a binary state (0 or 1). Formally, such systems are described as (1+1)-dimensional cellular automata, where the first dimension represents discrete spatial extent and the second corresponds to discrete temporal evolution. When restricted to nearest-neighbor interactions, the neighborhood of a cell comprises itself and its immediate left and right neighbors, forming cellular automata with a three-cell neighborhood. This type of automaton is referred to as an elementary cellular automaton. 

The local rules governing 1-dimensional cellular automata with a three-cell neighborhood can be fully specified by an 8-bit binary number. Each such binary number corresponds to a unique rule identifier expressed as a decimal value between 0 and 255, resulting in 256 distinct possible rules. This arises because the next state of a cell depends on its own state and those of its two immediate neighbors—a 3-cell configuration with 2 possible states per cell, yielding $2^3$ neighborhood patterns. Each pattern maps to one of 2 possible output states, leading to $2^{2^3}$ = 256 total rules.

\begin{figure}[ht]
  \centering
    \includegraphics[width=1\columnwidth]{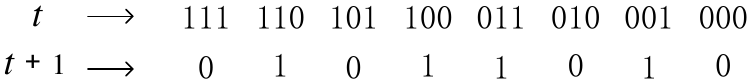}
    \caption{An example of the local rule for the time evolution of one-dimensional Wolfram automaton. The numerical representation of the rule 90. According to equation (\ref{eqs:rule_EV}), this gives rule 90.}
   \label{fig:rule_N_90_1}
\end{figure}

The output values $a_i$ for each input condition of the Wolfram automaton are 
arranged in descending order of the corresponding neighborhood state index (i.e.,
$a_7$ to $a_0$), forming an 8-bit binary number. The rule number $R$ is computed as the decimal equivalent of this binary representation:
\begin{equation}
R=\sum_{i=0}^{7}{2^ia_i},
\label{eqs:rule_EV}
\end{equation}
where $a_i$ denotes the output state for the $i$-th neighborhood configuration. For example, Fig.~\ref{fig:rule_N_90_1} shows the numerical representation of the Wolfram rule 90 (hereinafter rule 90), where the binary sequence 01011010 yields $R=2^6+2^4+2^3+ 2^1=90$.

The density of the Wolfram automata is defined as the ensemble average of the fraction of active sites(sites value of 1) at time step $t$:
\begin{equation}
 \rho(t)=\big\langle \frac{1}{L}\sum_{i} s_i(t)\big\rangle\,,
\label{eqs:orderp}
 \end{equation}
where $L$ is the lattice size, $s_i(t) \in \{0,1\}$ is the state of site $i$ at time step $t$, and $\left\langle \cdot \right\rangle$ denotes averaging over independent initial conditions. After the long time evolution, $\rho (t)$ converges to a stationary value $\rho_{stat}$. In the thermodynamic limit ($L\to\infty$, $t\to\infty$), this stationary density defines the asymptotic density $\rho_{\infty}$.

\section{Numerical computation of Wolfram automata \label{sec:Wolfram numerical}}

We first employ Monte Carlo simulation methods to study the Wolfram automata, leveraging randomly generated initial configurations to analyze fractal structures and self-organization phenomena. This approach specifically investigates the relationship between the asymptotic density of evolved states and the disordered initial density of the automaton. Subsequently, we utilize numerical computation to explore the evolution mechanisms of cellular automata configurations under predefined Wolfram rules.

\begin{figure*}[t]
\setlength{\tabcolsep}{0pt}
\centering
\begin{tabular}{cc}
\centering
\subfigure{
\begin{minipage}[t]{0.5\linewidth}
\includegraphics[width=8.2cm]{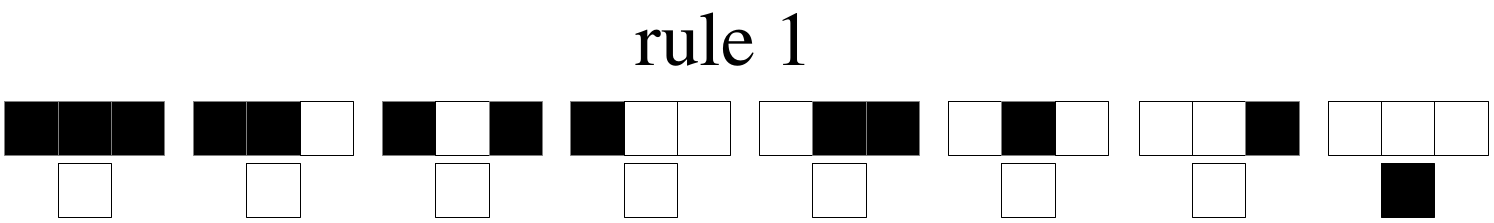}
\end{minipage}%
}&
\subfigure{
\begin{minipage}[t]{0.5\linewidth}
\includegraphics[width=8.2cm]{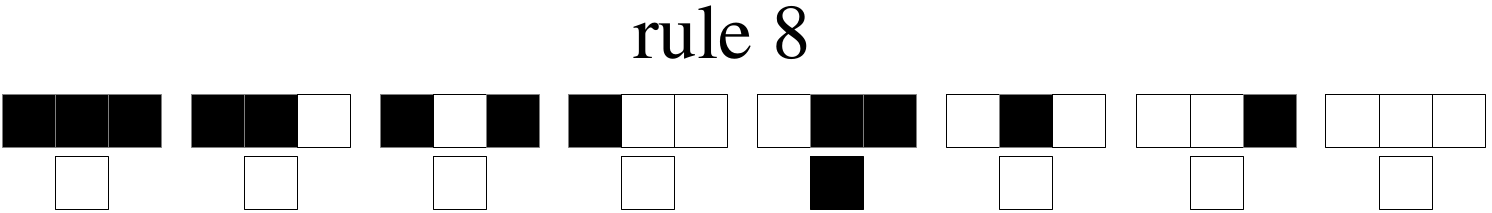}
\end{minipage}%
}\\
\subfigure{
\begin{minipage}[t]{0.5\linewidth}
\includegraphics[width=8.2cm]{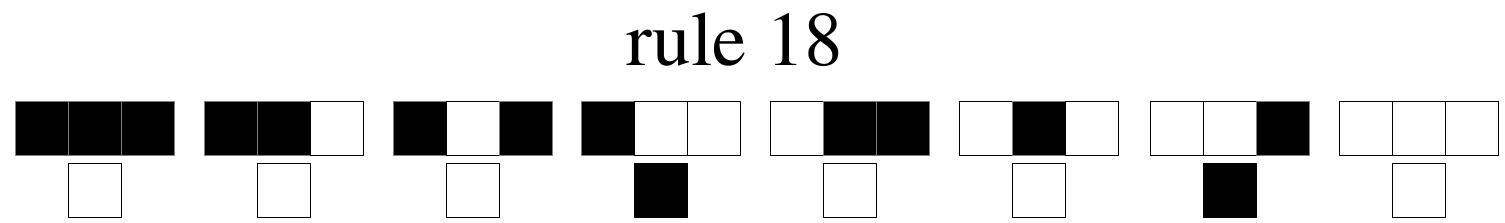}
\end{minipage}%
}&
\subfigure{
\begin{minipage}[t]{0.5\linewidth}
\includegraphics[width=8.2cm]{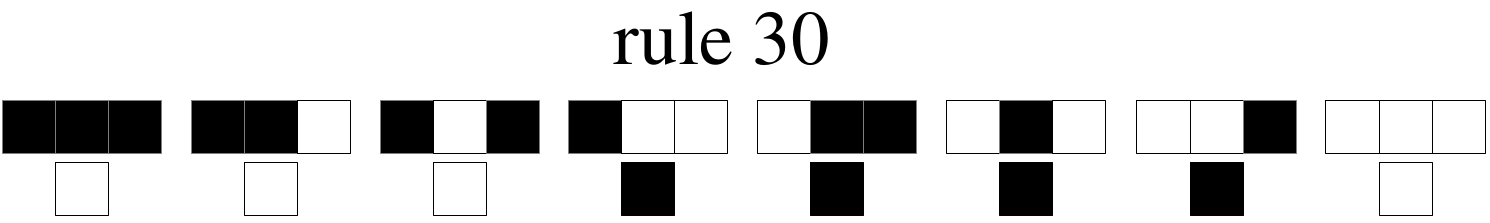}
\end{minipage}%
}\\
\subfigure{
\begin{minipage}[t]{0.5\linewidth}
\includegraphics[width=8.2cm]{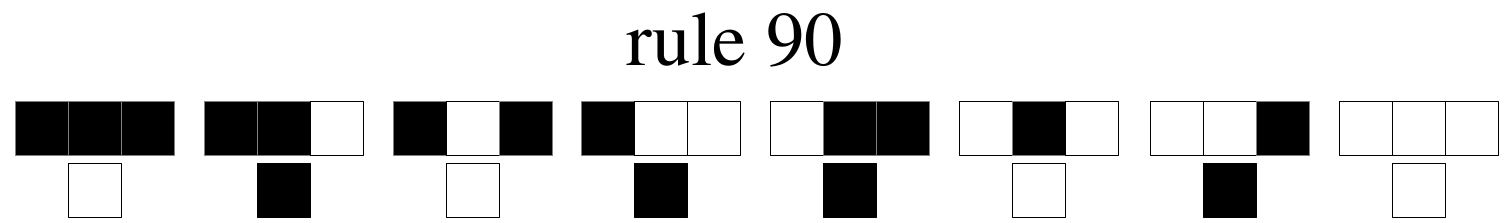}
\end{minipage}%
}&
\subfigure{
\begin{minipage}[t]{0.5\linewidth}
\includegraphics[width=8.2cm]{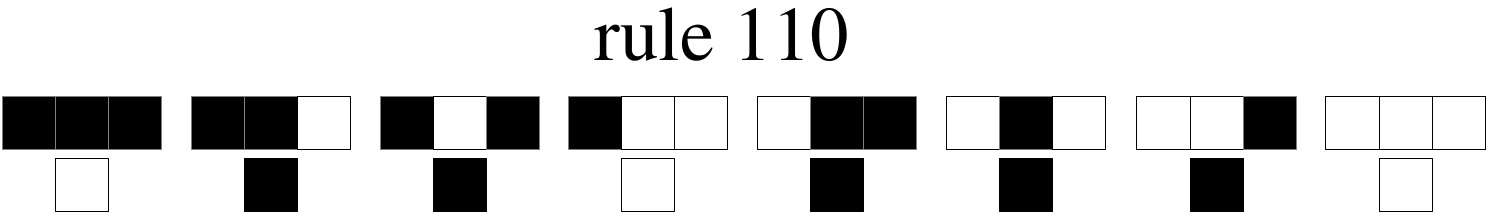}
\end{minipage}%
}\\
\subfigure{
\begin{minipage}[t]{0.5\linewidth}
\includegraphics[width=8.2cm]{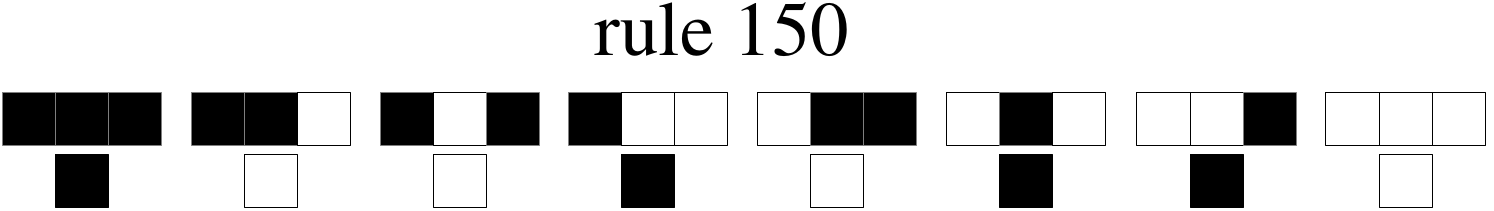}
\end{minipage}%
}&
\subfigure{
\begin{minipage}[t]{0.5\linewidth}
\includegraphics[width=8.2cm]{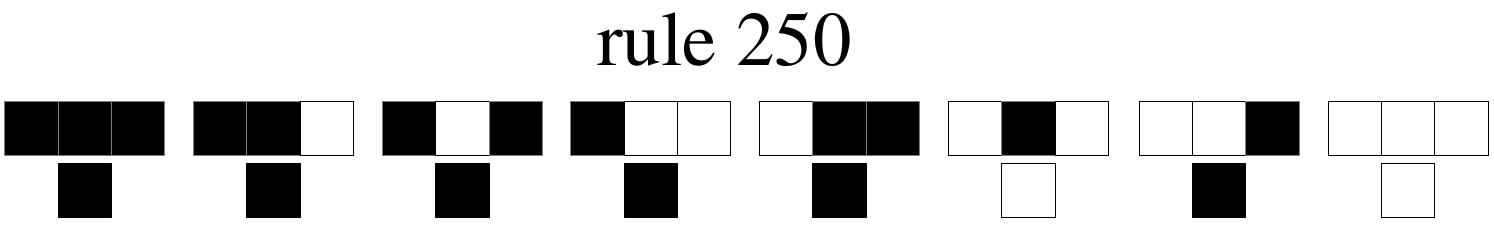}
\end{minipage}%
}\\
\end{tabular}
\caption{A black-and-white representation of the Wolfram rule. Black color represents
sites with value 1 and white represents empty sites.}
\label{fig:Black_white_Wolfram_rule}
\end{figure*}

\begin{figure*}[htbp]
\setlength{\tabcolsep}{1.2pt}
\centering
\begin{tabular}{ccc}
\includegraphics[width=0.45\columnwidth]{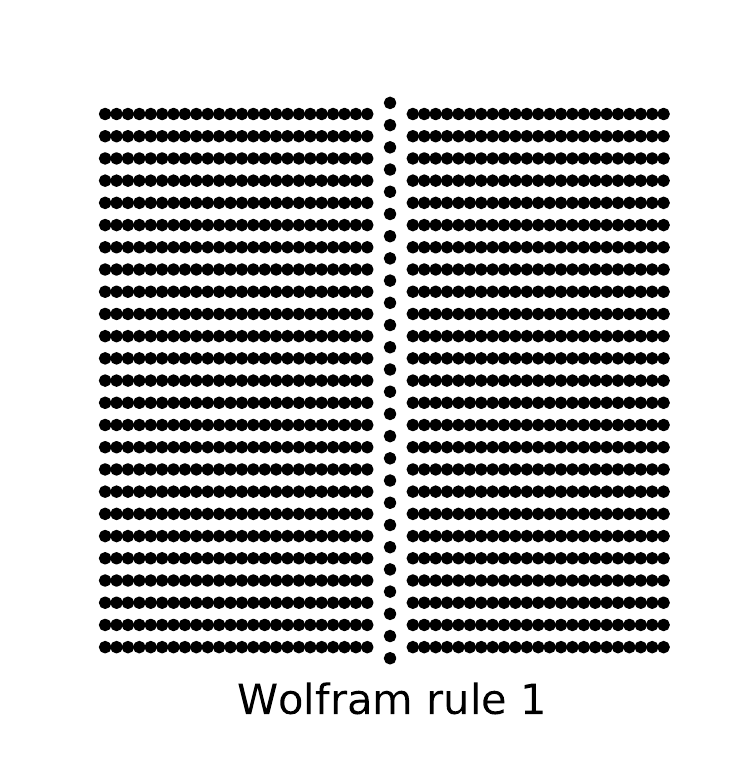}&
\includegraphics[width=0.45\columnwidth]{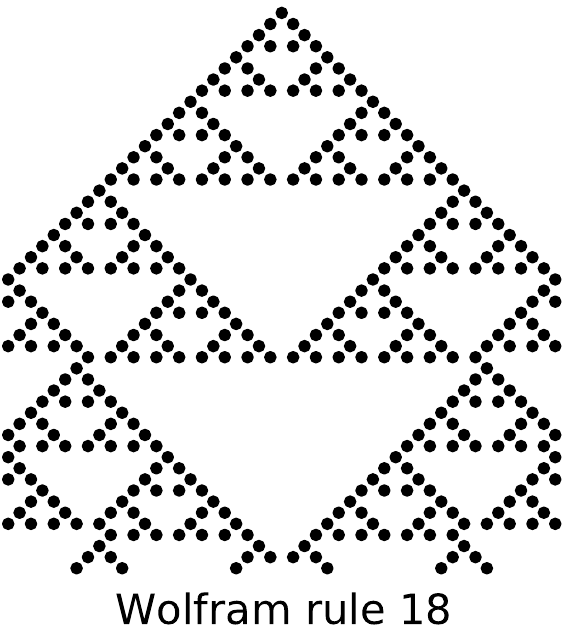}&
\includegraphics[width=0.45\columnwidth]{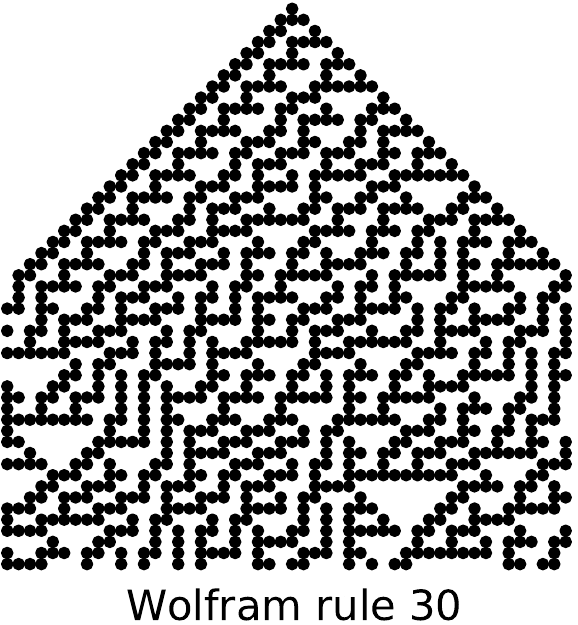}\\
\includegraphics[width=0.45\columnwidth]{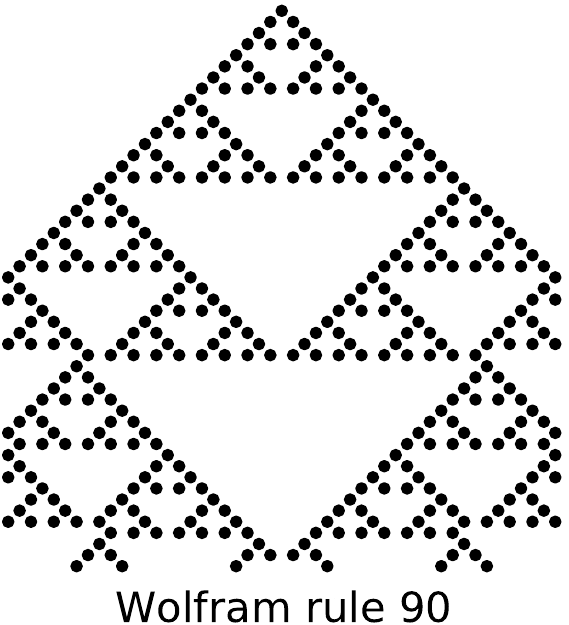}&
\includegraphics[width=0.45\columnwidth]{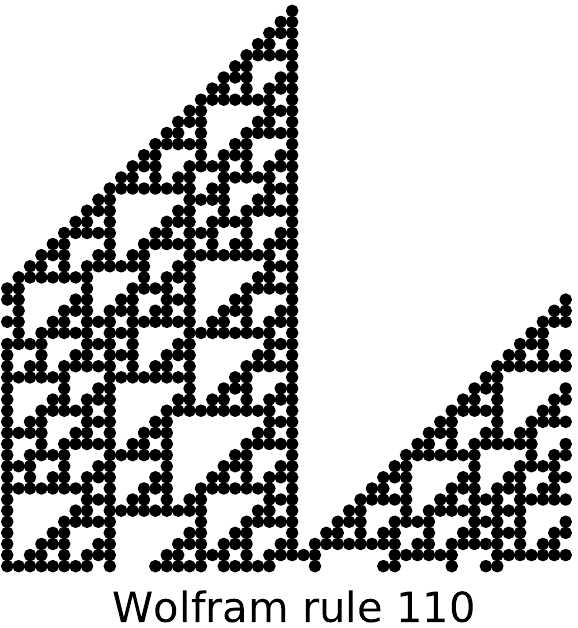}&
\includegraphics[width=0.45\columnwidth]{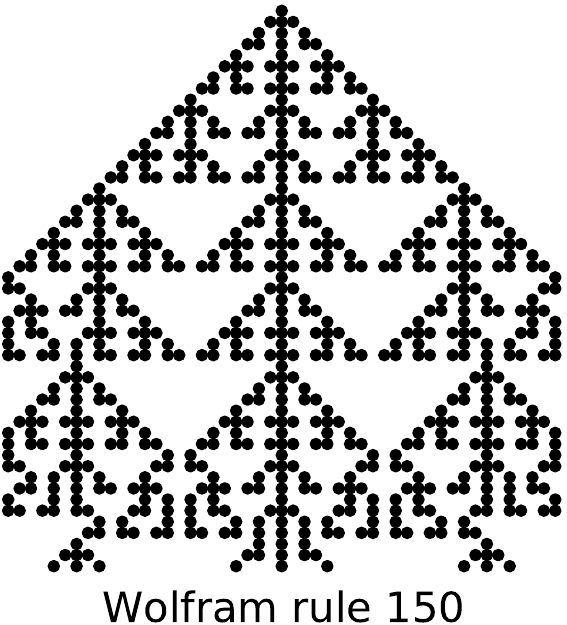}\\
\end{tabular}
\caption{Evolution of the configurations of the Wolfram automaton under the initial state with a single site value of 1. Size $L=50$, time steps $t=50$.}
\label{fig:Black_white_Wolfram_rule_single}
\end{figure*}

The local rules governing Wolfram automata can be formalized as Boolean functions acting on a central site and its nearest neighbors. Fig.~\ref{fig:Black_white_Wolfram_rule} shows a black-and-white color representation of the Wolfram rule, where black color represents sites with value 1 and white represents sites with value 0. Each rule is defined by a unique Boolean expression. For example, the rule 90 is equivalent to $s_{i}(t+1)$ = $s_{i-1}(t) \oplus s_{i+1}(t)$, the rule 18 to $s_{i}(t+1)$ = $\neg s_i(t) \land (s_{i-1}(t) \oplus s_{i+1}(t))$, the rule 30 to $s_{i}(t+1)$ = $s_{i-1}(t) \oplus (s_i(t) \lor s_{i+1}(t))$, and the rule 110 to $s_{i}(t+1)$ = $(s_i(t) \land (\neg s_{i-1}(t))) \lor (s_i(t) \oplus s_{i+1}(t))$. Here, $s_i(t)$ denotes the state of site $i$ at time step $t$, while $s_{i-1}(t)$ and $ s_{i+1}(t)$ represent its left and right neighbors, respectively. Boolean operators include AND ($\land$), OR ($\lor$), NOT ($\neg$), and XOR ($\oplus$). Compared to enumerating all 256 rule conditions, these simplified Boolean expressions significantly enhance computational efficiency in simulations by reducing logical redundancy and enabling optimized code implementation.

\begin{figure*}[htbp]
\setlength{\tabcolsep}{0pt}
\centering
\begin{tabular}{ccc}
\centering
\subfigure{
\begin{minipage}[t]{0.33\linewidth}
\includegraphics[width=5.6cm]{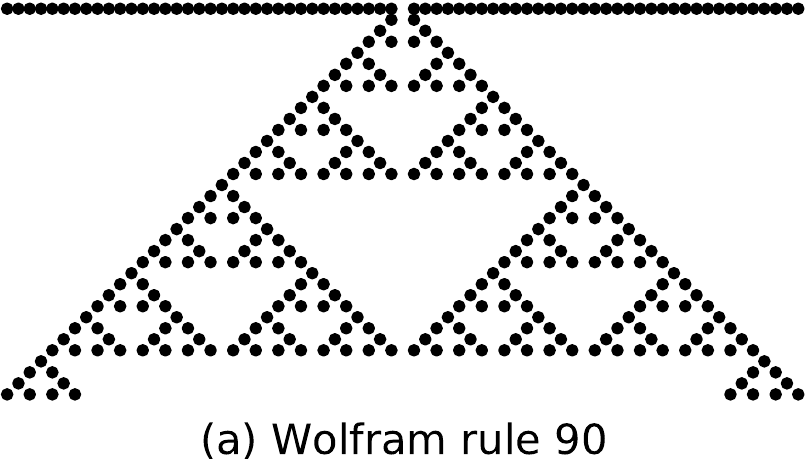}
\end{minipage}%
}&
\subfigure{
\begin{minipage}[t]{0.33\linewidth}
\includegraphics[width=5.6cm]{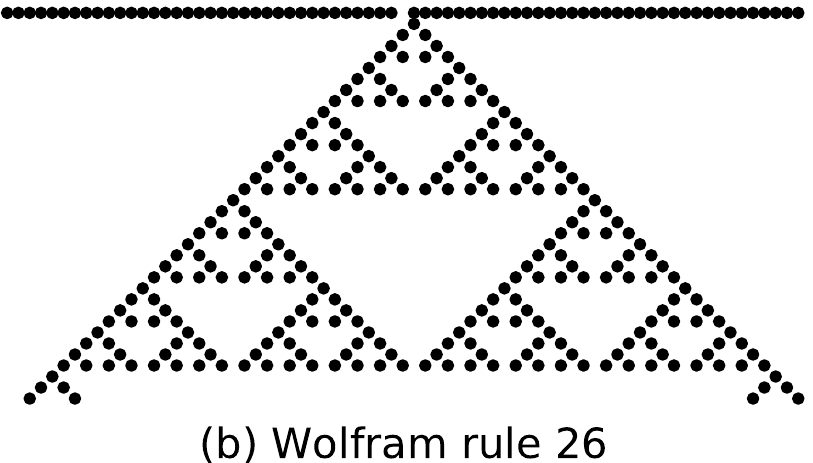}
\end{minipage}%
}&
\subfigure{
\begin{minipage}[t]{0.33\linewidth}
\includegraphics[width=5.6cm]{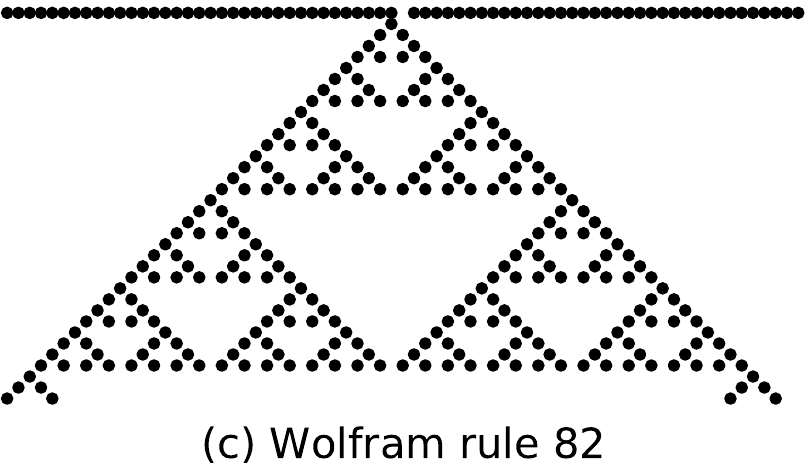}
\end{minipage}%
}\\
\subfigure{
\begin{minipage}[t]{0.33\linewidth}
\includegraphics[width=5.6cm]{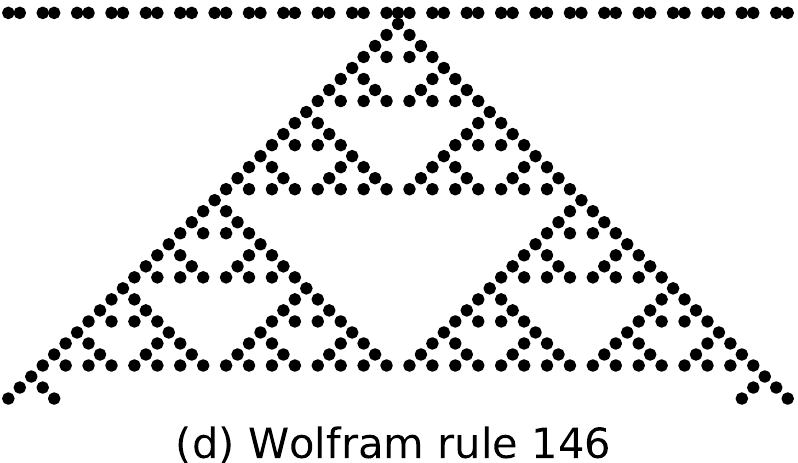}
\end{minipage}%
}&
\subfigure{
\begin{minipage}[t]{0.33\linewidth}
\includegraphics[width=5.6cm]{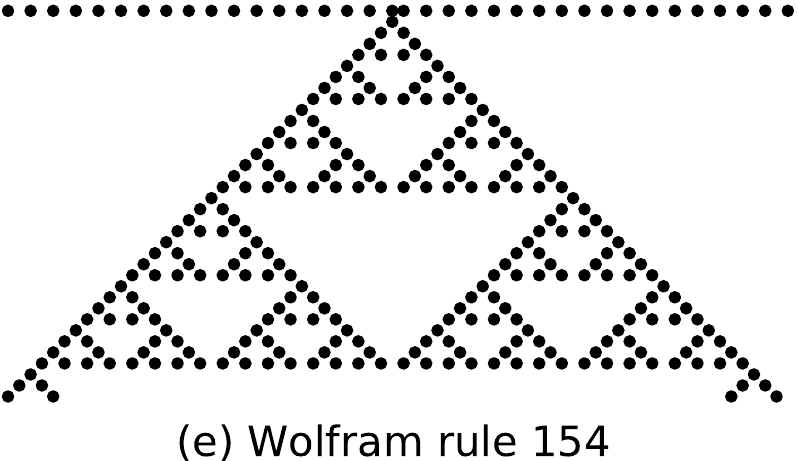}
\end{minipage}%
}&
\subfigure{
\begin{minipage}[t]{0.33\linewidth}
\includegraphics[width=5.6cm]{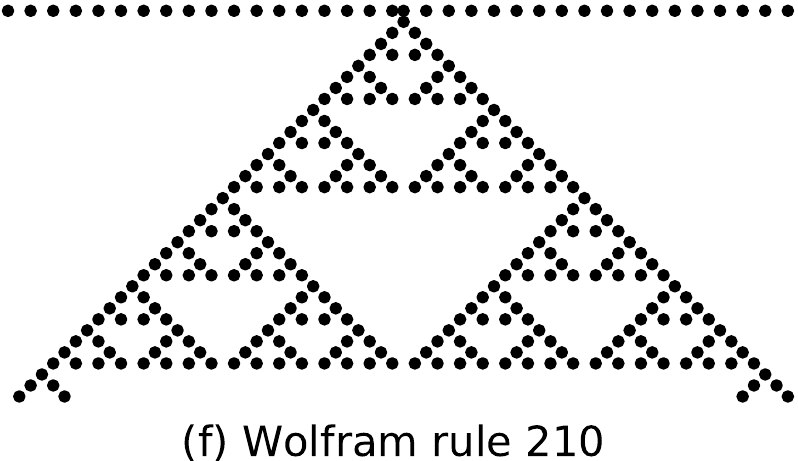}
\end{minipage}%
}
\end{tabular}
\caption{Evolution of the configurations of (a) the rule 90, (b) the rule 26, and (c) the rule 82 under an initial state where the central site is 0 and all other sites are 1. Evolution of the configurations of (d) the rule 146 under the initial state ``\dots 0, 1, 1, 0, 1, 1, 0, 1, 1, 1, 0, 1, 1, 0, 1, 1 \dots'', (e) the rule 154 under the initial state ``\dots 0, 1, 0, 1, 0, 1, 1, 0, 1, 0, 1 \dots'', and (f) the rule 210 under the initial state ``\dots 0, 1, 0, 1, 0, 1, 1, 0, 1, 0, 1 \dots''.}
\label{fig:rule_90_26_82_single_0_1}
\end{figure*}

 As shown in Fig.~\ref{fig:Black_white_Wolfram_rule_single}, the cluster diagram of the Wolfram automaton evolves from an initial state with a single site value of 1. Starting from such simple initial configurations, Wolfram automata either evolve converge to homogeneous states or generate self-similar fractal patterns. For example, under the initial state with a single site value of 1, the rules 18 and 90 both produce the Sierpinski triangle pattern with a fractal dimension of $\log_2 3 \approx 1.585$. Similarly, the rule 150 exhibits self-similarity with a fractal dimension of 1.69.
 
Numerical analysis reveals that the emergent Sierpinski triangle structure in the rules 18 and 90 depends solely on two specific rule conditions: neighborhood state \texttt{100$\rightarrow$1} and \texttt{001$\rightarrow$1}. Notably, the rule 90 additionally satisfies conditions such as \texttt{110$\rightarrow$1} and \texttt{011$\rightarrow$1}, but these do not influence the Sierpinski pattern under the specified initial state. This property extends to other rules satisfying the same core conditions (\texttt{100$\rightarrow$1} and \texttt{001$\rightarrow$1}), resulting in eight distinct Wolfram rules (including the rules 26, 82, 146, 154, 210, and 218) that generate Sierpinski triangles when evolved from configurations containing only one active site at any time step.

Further investigations under modified initial conditions highlight nuanced behavioral differences.
For the rule 90 evolving from an initial state where the central site is 0 and all others are 1, the state transition at $t=1$ depends exclusively on \texttt{110$\rightarrow$1} and \texttt{011$\rightarrow$1}. This results in the activation of two central sites at this time step and ultimately forming a Sierpinski triangle with vertices removed (Fig.~\ref{fig:rule_90_26_82_single_0_1}(a)). Under identical initial conditions, the transition
of the rule 26 at $t=1$ depends solely on \texttt{011$\rightarrow$1}, while the rule 82 depends solely on \texttt{110$\rightarrow$1}. Both rules yielding only one active site at this time step, and yet subsequently evolve into full Sierpinski triangles (Figs.~\ref{fig:rule_90_26_82_single_0_1}(b)–(c)). For specialized initial states (e.g., alternating sequences with isolated active sites), the rules 146, 154, and 210 each activate only one site at $t=1$ due to distinct rule dependencies (\texttt{111$\rightarrow$1} for the rule 146, \texttt{011$\rightarrow$1} for the rule 154, and \texttt{110$\rightarrow$1} for the rule 210), yet all subsequently generate complete Sierpinski triangles(Figs.~\ref{fig:rule_90_26_82_single_0_1}(d)–(f)).

\begin{figure}[htbp]
\setlength{\tabcolsep}{1.2pt}
\centering
\begin{tabular}{cc}
\subfigure{
\begin{minipage}[t]{0.49\linewidth}
\includegraphics[width=4.3cm]{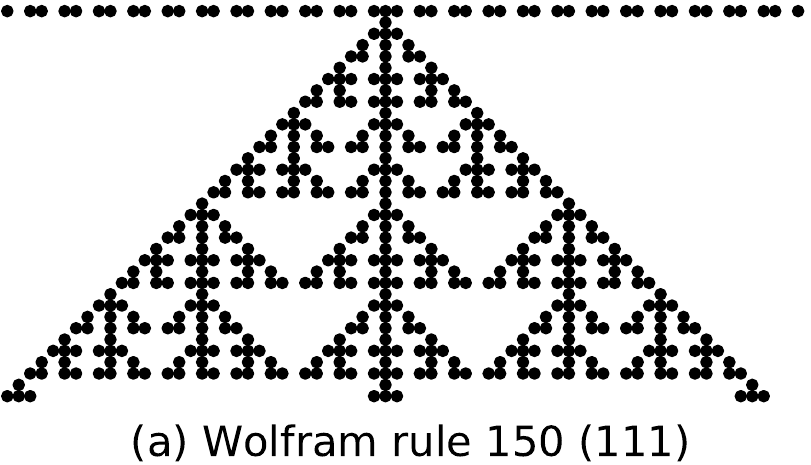}
\end{minipage}%
}&
\subfigure{
\begin{minipage}[t]{0.49\linewidth}
\includegraphics[width=4.3cm]{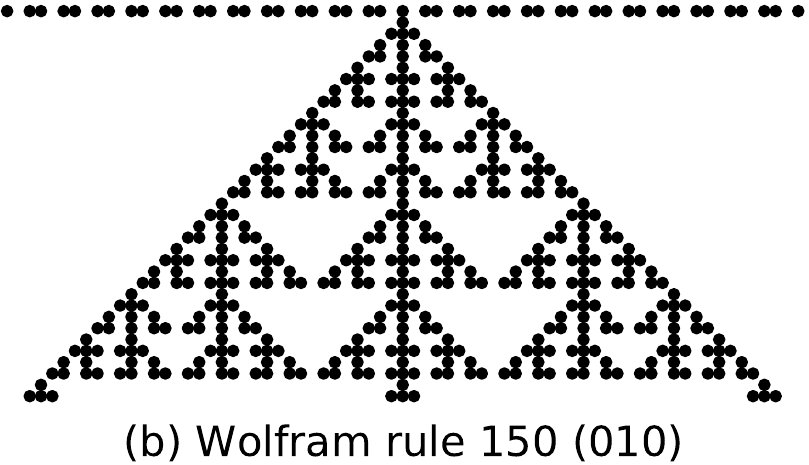}
\end{minipage}%
}
\end{tabular}
\caption{Evolution of the configurations of (a) the rule 150 under the initial state ``\dots 1, 1, 0, 1, 1, 0, \textbf{1, 1, 1}, 0, 1, 1, 0, 1, 1 \dots'', and (b) the rule 150 under the initial state ``\dots 1, 1, 0, 1, 1\textbf{, 0, 1, 0}, 1, 1, 0, 1, 1 \dots''.}
\label{fig:rule_150_111_010}
\end{figure}
Under the initial state with a single site value of 1, the rule 150 generates a self-similar pattern with a fractal dimension of 1.69 \cite{wolfram1983statistical}. Further numerical analysis reveals that if a configuration at any time step $t$, contains exactly one active site, the rule 150's subsequent evolution will consistently produce self-similar patterns with the same fractal dimension of 1.69. This property holds across diverse initial states, for the initial state ``\dots 1, 1, 0, 1, 1, 0, 1, 1, 1, 0, 1, 1, 0, 1, 1 \dots'', the state transition at $t=1$ depends solely on the rule condition \texttt{111$\rightarrow$1}, resulting in a single active site; for the initial state ``\dots 1, 1, 0, 1, 1, 0, 1, 0, 1, 1, 0, 1, 1 \dots'', the transition at $t=1$ relies exclusively on \texttt{010$\rightarrow$1}, also yielding one active site. Despite these distinct initial configurations and localized rule dependencies, the rule 150's long-term evolution converges to identical self-similar structures with fractal dimension 1.69, as shown in Fig.~\ref{fig:rule_150_111_010}.

Grassberger and Wolfram conducted theoretical analyses primarily focused on complex Wolfram rules \cite{wolfram1983statistical,grassberger1983new}, such as the rules 18, 90 and 182, which exhibit well-defined asymptotic density $\rho_\infty$. This asymptotic density represents the long-term stationary value $\rho_{stat}$ of the time-dependent density $\rho$(t), is typically independent of the initial state density. However, under specific initial conditions, for example, a configuration with only a single site active, the rules 18 and 90 generate Sierpinski triangle patterns. This implies that their asymptotic density displays periodic oscillations rather than converging to a constant value. The observed divergence from their established $\rho_\infty$ motivates further investigation into whether the asymptotic behavior of complex Wolfram rules under very low initial densities aligns with their behavior under disordered initial states. Notably, the steady-state density of many Wolfram rules are known to depend on initial conditions, which prompts a systematic exploration of the relationship between initial density and long-term statistical behavior. In this work, we employ numerical simulation and analysis to study the temporal evolution of configurations in these automata, aiming to characterize their dynamic properties.

\begin{figure}[htbp]
\setlength{\tabcolsep}{0pt}
\centering
\begin{tabular}{cc}
\includegraphics[width=0.49\columnwidth]{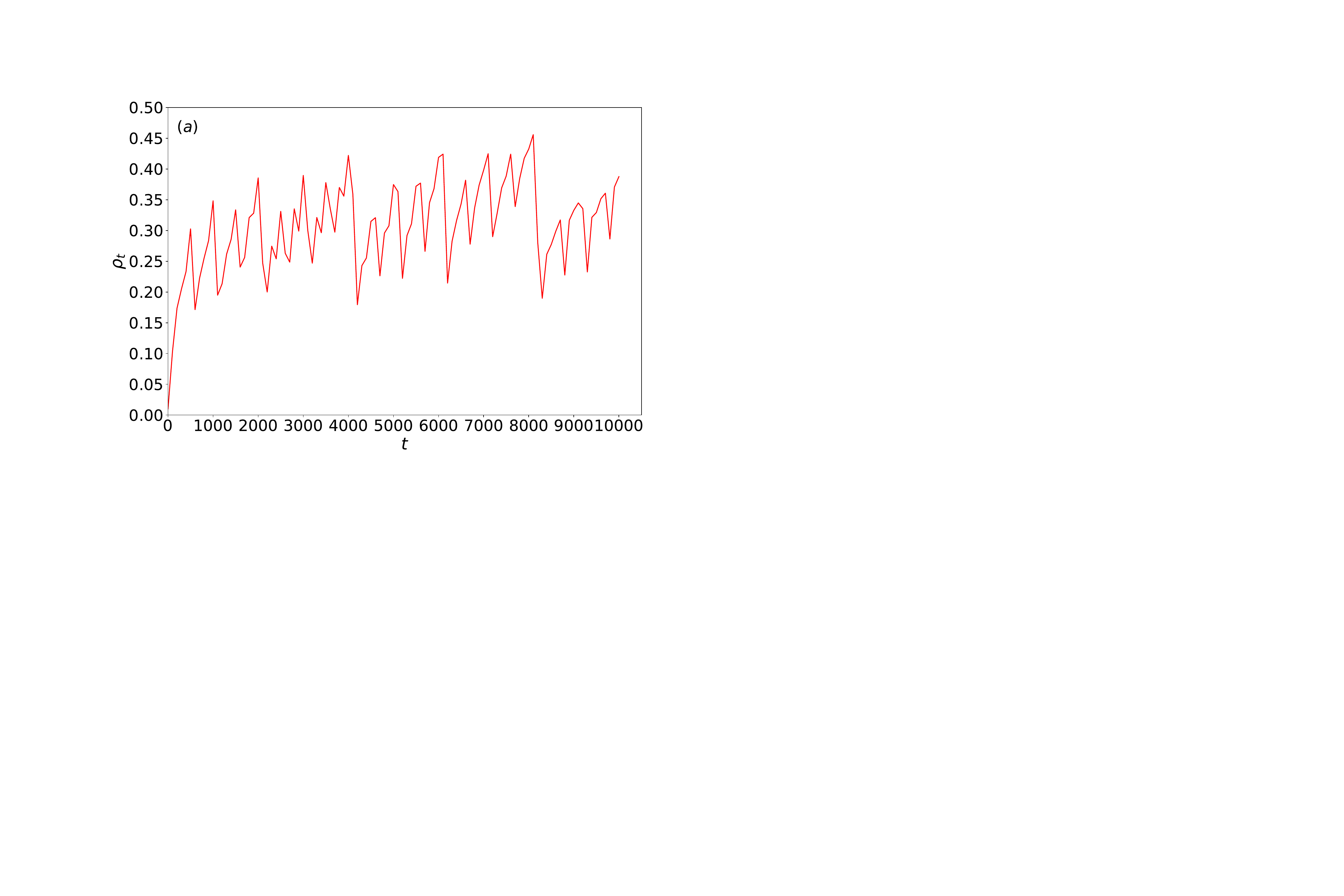} &
\includegraphics[width=0.49\columnwidth]{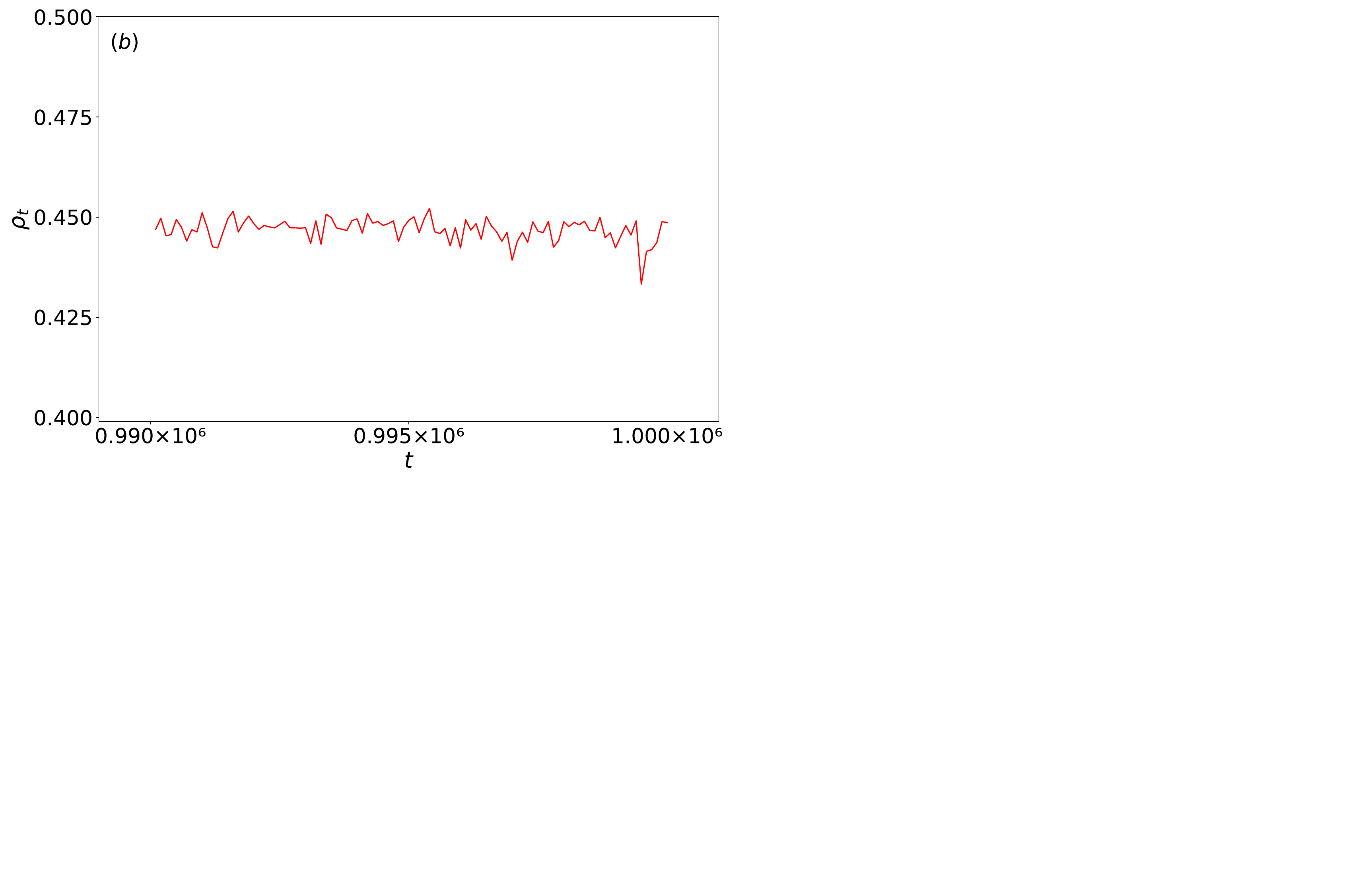} \\
\includegraphics[width=0.49\columnwidth]{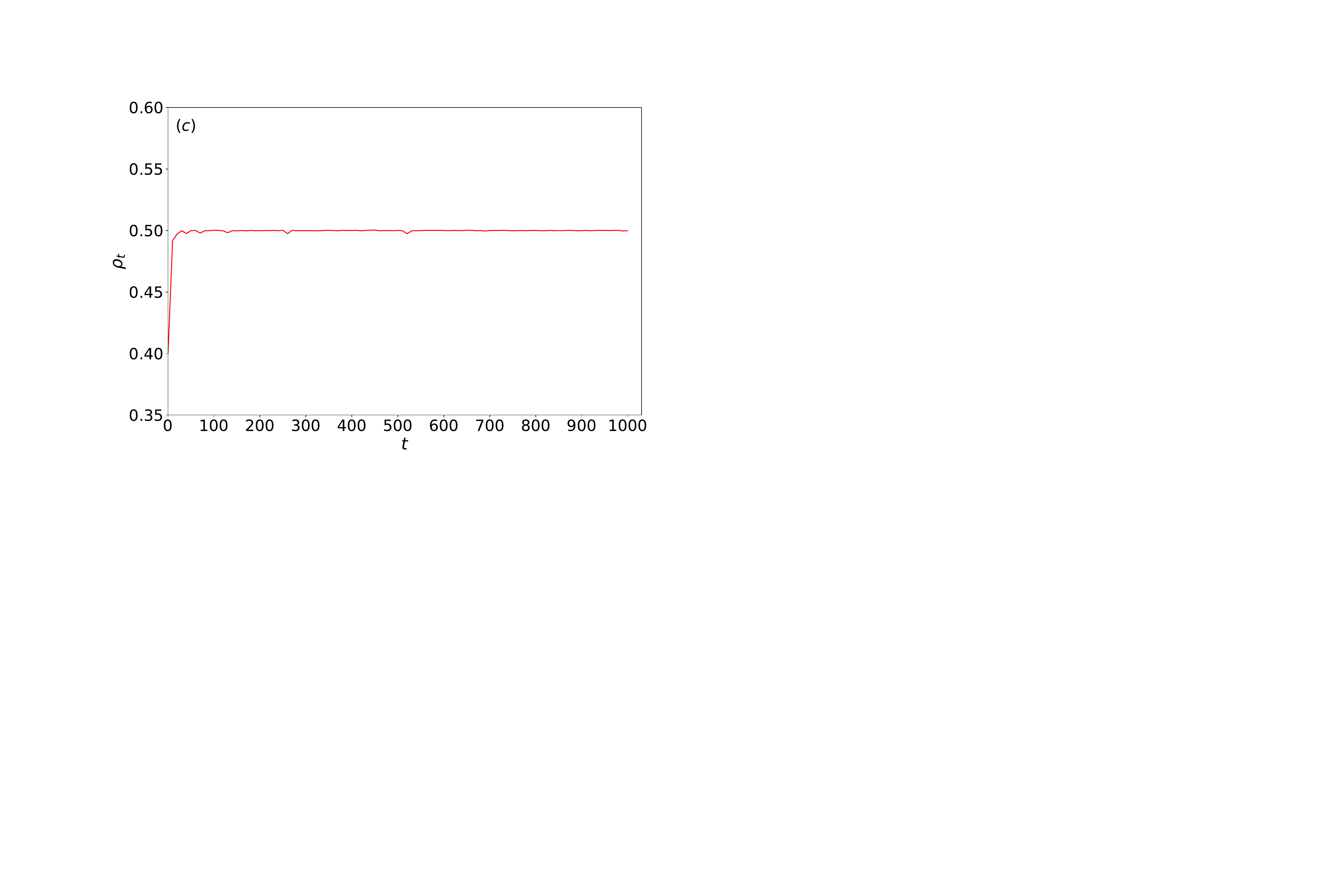}&
\includegraphics[width=0.49\columnwidth]{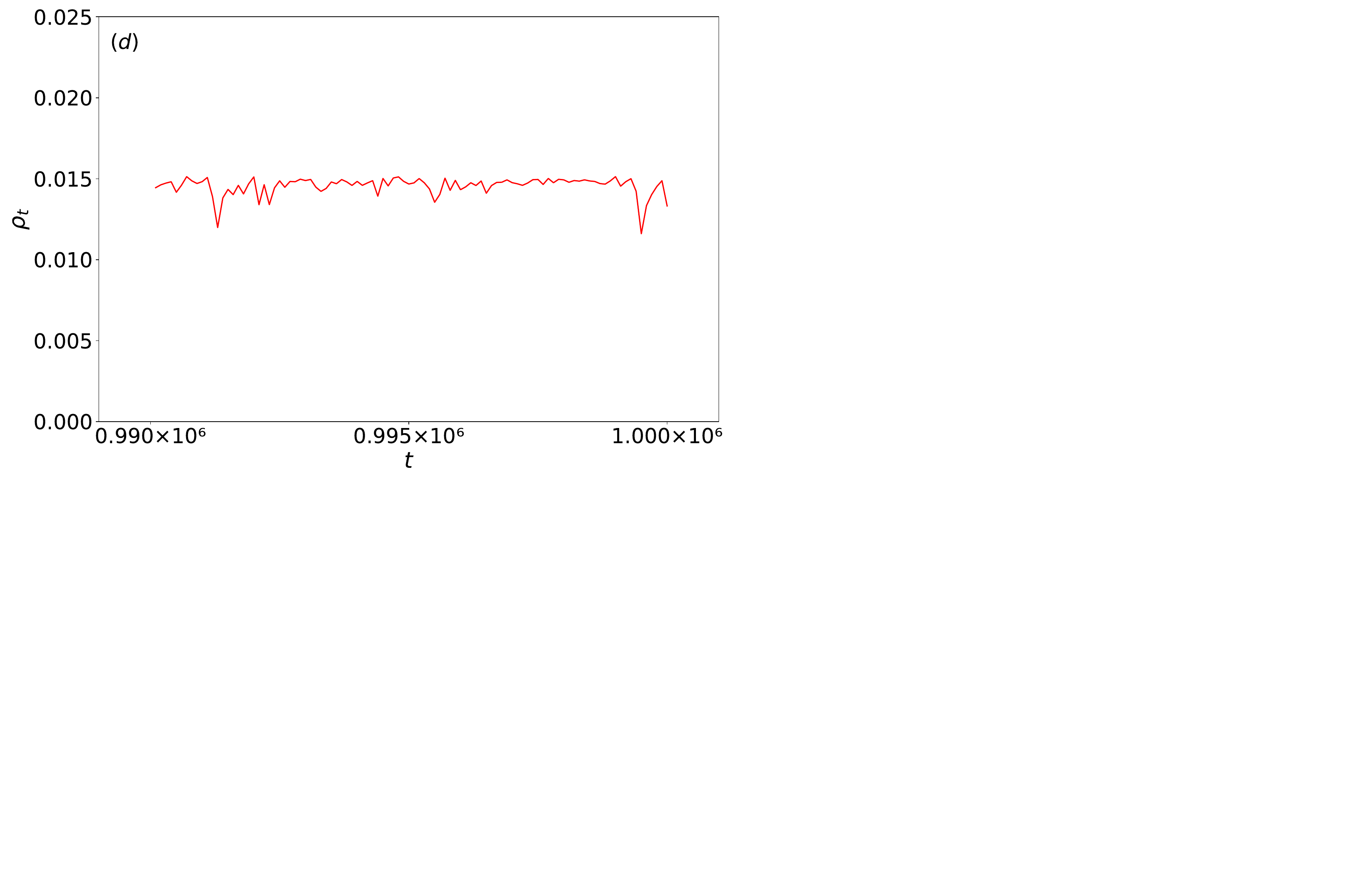}\\
\includegraphics[width=0.49\columnwidth]{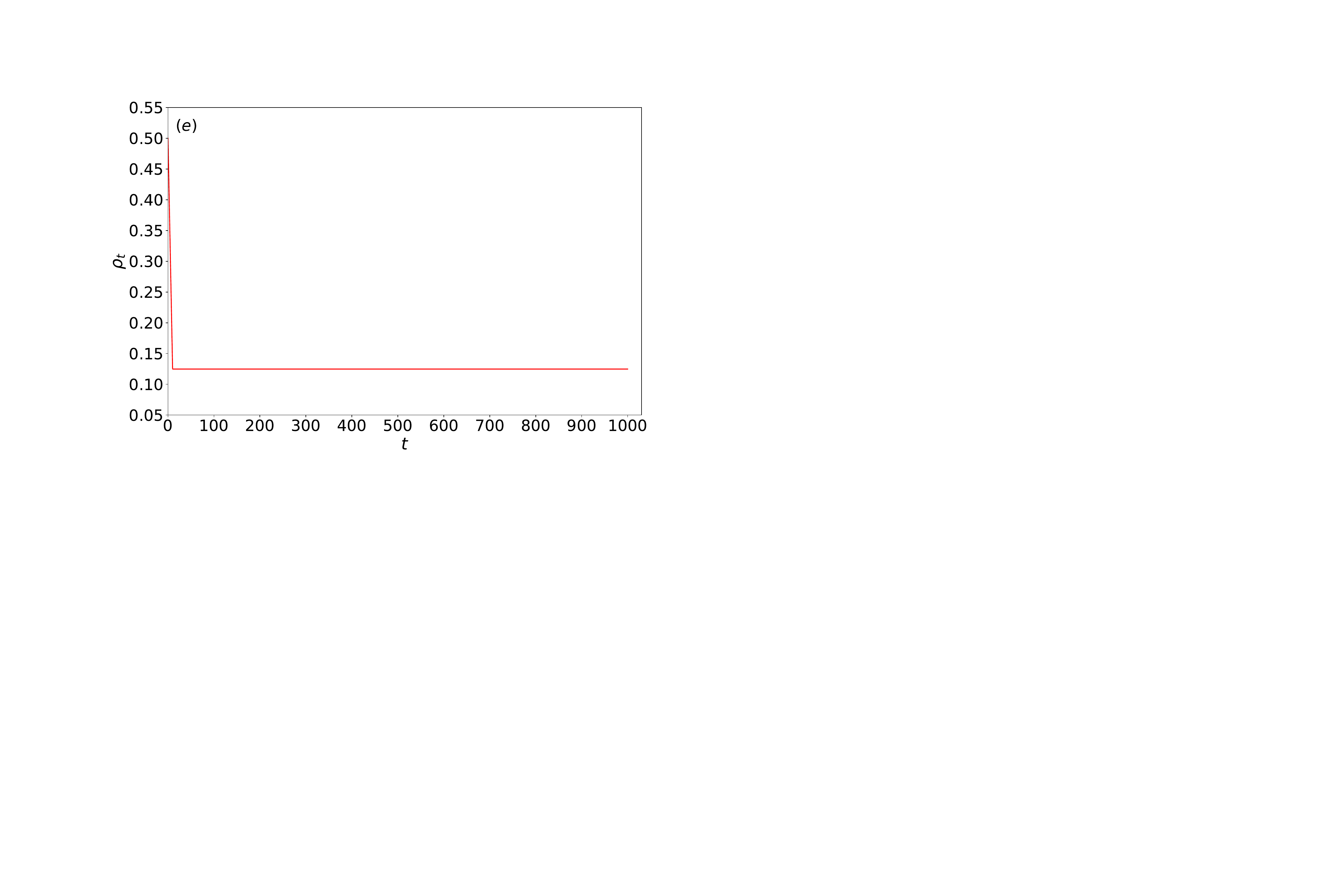} &
\includegraphics[width=0.49\columnwidth]{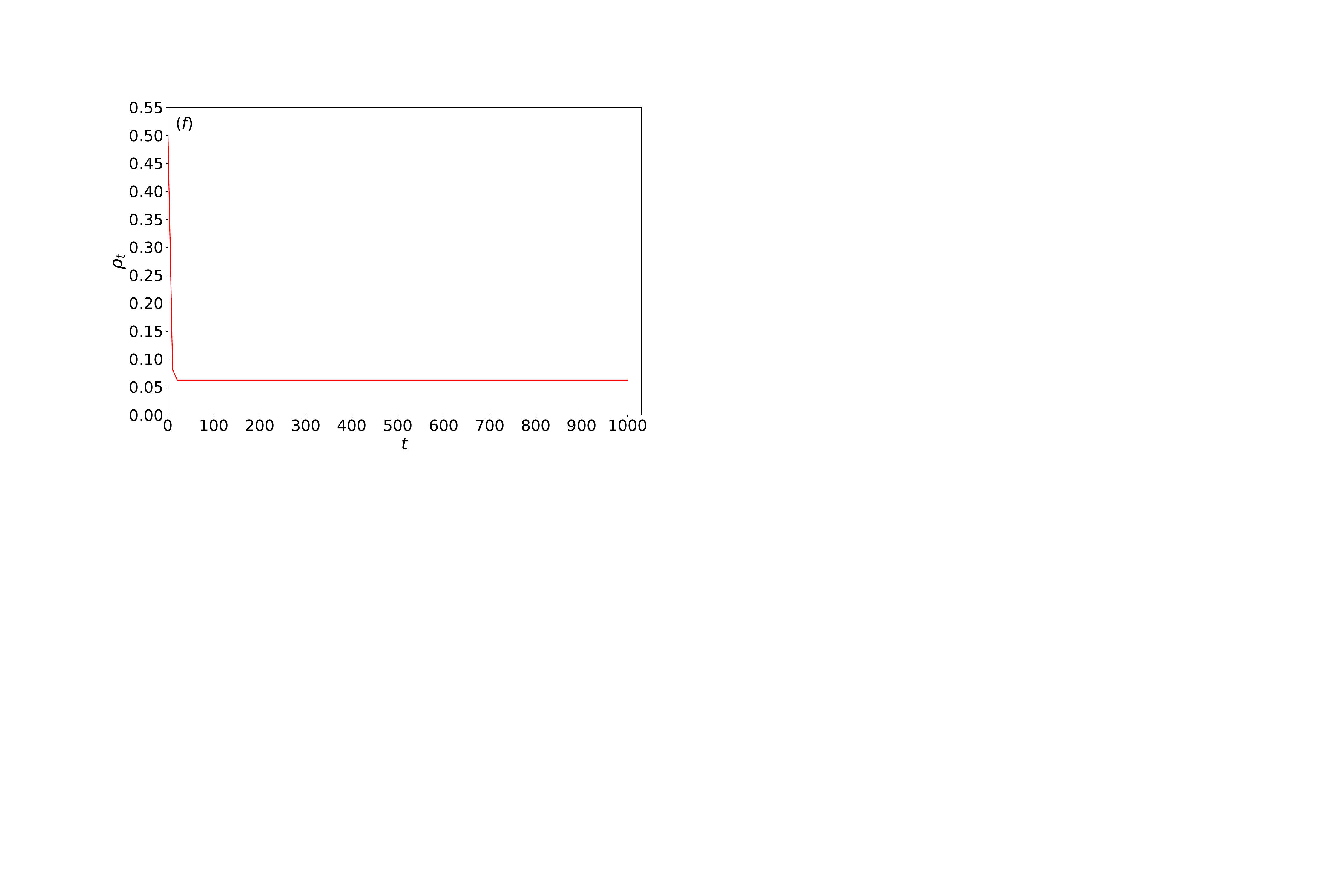}\\
\end{tabular}
\caption{The temporal evolution of density $\rho_t$ for various Wolfram rules under different initial densities. (a) $\rho_t$ for the rule 90 with initial density $\rho _0$ = 0.01, system size $L$ = 1000, and simulation steps $t$ = $10^4$. (b) The same rule 90 system under identical initial conditions, but during the last $10^4$ steps of an extended simulation ($10^6$). (c) The rule 90 with $\rho _0$ = 0.4, $L$ = 1000, and $t$ = 1000. (d) The rule 18 with $\rho _0$ = 0.99, $L$ = 1000, observed during the last $10^4$ steps of $10^6$. (e) The rule 2 with $\rho _0$ = 0.5, $L$ = 1000, and $t$ = 1000. (f) The rule 36 under identical initial conditions to the rule 2. For panels (a), (b), and (d), $\rho _t$ is computed as the moving average over configurations from $t-$99 to $t$, capturing longer-term statistical trends. For panels (c), (e), and (f), $\rho _t$ uses a shorter averaging window from $t-$9 to $t$, emphasizing transient dynamics.}
\label{fig:rule_90_18_2_36_MC_t}
\end{figure}

Monte Carlo simulations of (1+1)-dimensional Wolfram automata are conducted according to predefined Wolfram rules under periodic boundary conditions. When starting from a disordered initial state where each site has an independent probability $p$, the initial density $\rho _0$ equals $p$. Fig.~\ref{fig:rule_90_18_2_36_MC_t} illustrates the temporal evolution of density $\rho_t$ for various Wolfram rules under different initial densities.

 As shown in Figs.~\ref{fig:rule_90_18_2_36_MC_t} (a) and (b), for the rule 90 with a very low initial density ($\rho _0$ = 0.01, $L$ = 1000), the density exhibits significant fluctuations over the first $t$ = $10^4$ time steps. During the last $10^4$ steps of a prolonged simulation ($t$ = $10^6$), $\rho_t$ tends to stabilize around approximately 0.45, though still displaying minor fluctuations. This stabilized value is notably lower than the theoretical asymptotic density $\rho_\infty = 0.5$. In contrast, for the rule 90 with a moderate initial density ($\rho _0$ = 0.4, $L$ = 1000, $t$ = 1000), as seen in Fig.~\ref{fig:rule_90_18_2_36_MC_t} (c), $\rho_t$ converges rapidly to the asymptotic value of 0.5 within 600 time steps and remains stable. 

For the rule 18 with a very high initial density ($\rho _0$ = 0.99, $L$ = 1000), Fig.~\ref{fig:rule_90_18_2_36_MC_t} (d) shows that during the last $10^4$ steps of $10^6$ steps simulation, $\rho_t$ oscillates around 0.0145 with small fluctuations, deviating significantly from its theoretical asymptotic density $\rho_\infty=0.25$.

In the cases of the rules 2 and 36 with initial density $\rho _0$ = 0.4 (Figs.~\ref{fig:rule_90_18_2_36_MC_t} (e) and (f)), both rules achieve stability quickly: the rule 2 stabilizes almost immediately, while the rule 36 converges within a few time steps.

 A key observation is that for complex rules like 18 and 90, under typical initial densities (e.g., $\rho _0$ = 0.4), $\rho_t$ stabilizes within a few hundred steps. However, under extreme initial densities (very low or very high), even prolonged evolution (e.g., $t$ = $10^6$) fails to achieve convergence to the theoretical asymptotic density, with $\rho_t$ exhibiting substantial fluctuations and persistent deviations from $\rho_\infty$. This behavior contradicts the theoretical expectation that asymptotic density should be independent of initial conditions for these automata. In contrast, simpler rules (such as the rules 2 and 36) exhibit rapid convergence and stable densities regardless of initial conditions.

\begin{figure}[htbp]
\setlength{\tabcolsep}{1.2pt}
\centering
\begin{tabular}{cc}
\includegraphics[width=0.493\columnwidth]{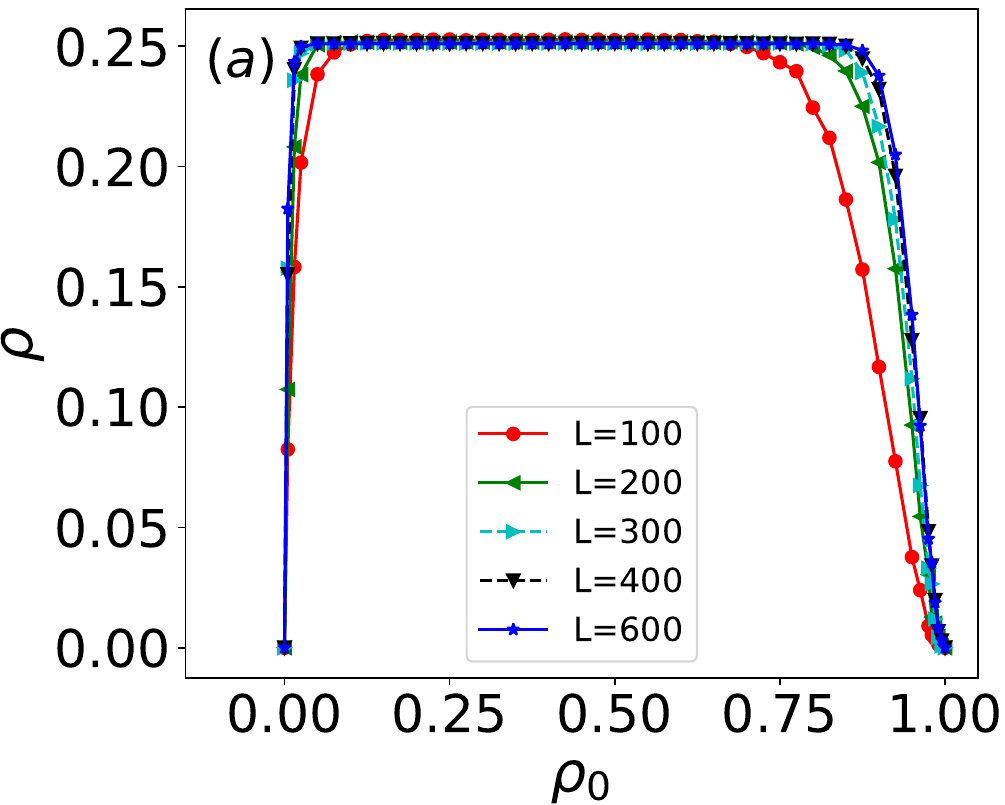}&
\includegraphics[width=0.481\columnwidth]{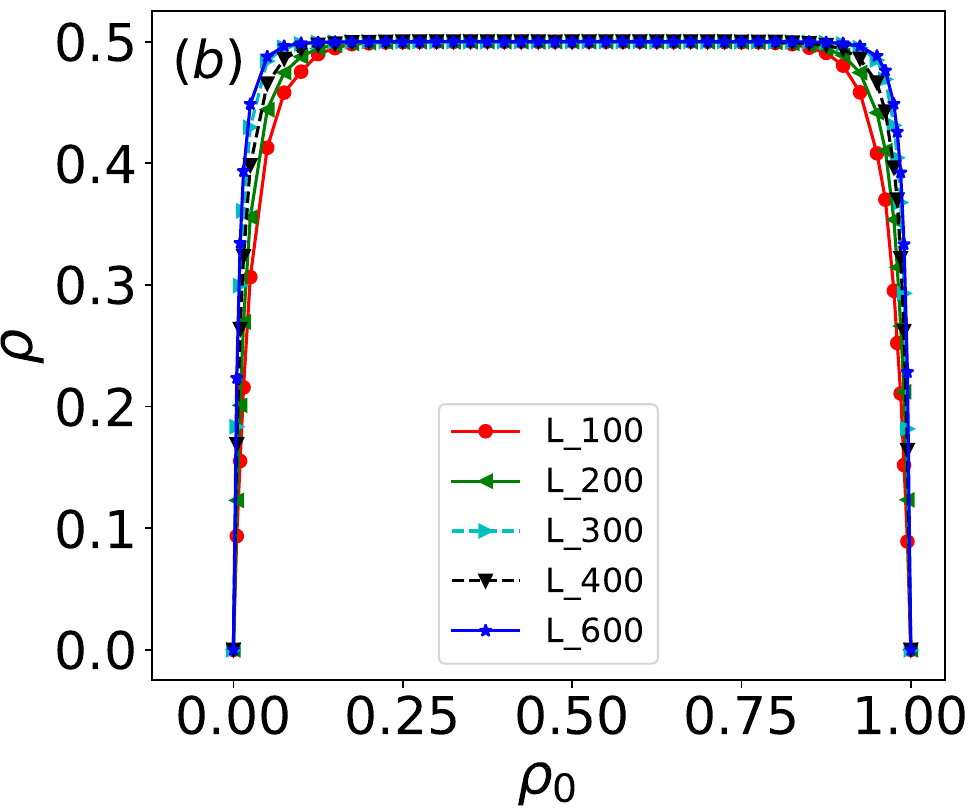}\\
\end{tabular}
\caption{Density $\rho$ of (a) the rule 18 and (b) the rule 90 under long-time evolution for disordered initial states with density $\rho _0$. where $L$ = 100, 200, 300, 400, 600; The corresponding time steps $t$ = 1000, 2000, 3000, 4000, 5000. }
\label{fig:Densities_Wolfram_rule_18_90_p=L_100_600}
\end{figure}

For the rules 18 and 90, simulations are run on arrays of sizes $L$ = 100, 200, 300, 400, and 600, and time steps $t$ = 1000, 2000, 3000, 4000, 5000, with the configurations from the last 200 time steps ($\Delta t$ = 200) being taken. For each initial density $\rho _0$, 1000 configurations are generated to calculate the density. As shown in the  Fig.~\ref{fig:Densities_Wolfram_rule_18_90_p=L_100_600}, the long-time evolved density of both rules under disordered initial states is presented. For most initial densities $\rho _0$, the rules 18 and 90 exhibit well-defined asymptotic densities: $\rho_\infty =0.25 \pm 0.002$ for the rule 18 and $\rho_\infty =0.5 \pm 0.002$ for the rule 90. However, under extreme initial densities (e.g., $\rho _0 \leq 0.01$ or $\rho _0\geq 0.99$), the evolved density significantly deviates from these asymptotic values.

At extremely low initial densities (characterized by very few active sites), the evolution of both rules depends primarily on the transition conditions \texttt{100$\rightarrow$1} and \texttt{001$\rightarrow$1}. In this regime, sites evolve nearly independently. When only a single site is active in the initial state, both rules generate Sierpinski triangle patterns, resulting in densities much lower than their asymptotic values. Under such sparse initial conditions, the evolution produces numerous isolated triangular structures with large empty regions, leading to low overall density. As the initial density increases, the evolved density converges toward the asymptotic value. Conversely, at extremely high initial densities, the first time step is dominated by the condition \texttt{111$\rightarrow$0}, causing a sharp drop in density. This results in configurations similar to those observed under very low initial densities. As the initial density decreases from this extreme, the evolved density again approaches the asymptotic value. For a fixed initial density, larger system sizes (increasing $L$)
yield densities closer to the asymptotic limit.

\begin{figure}[htbp]
\setlength{\tabcolsep}{1.2pt}
\centering
\begin{tabular}{c}
\includegraphics[width=0.49\columnwidth]{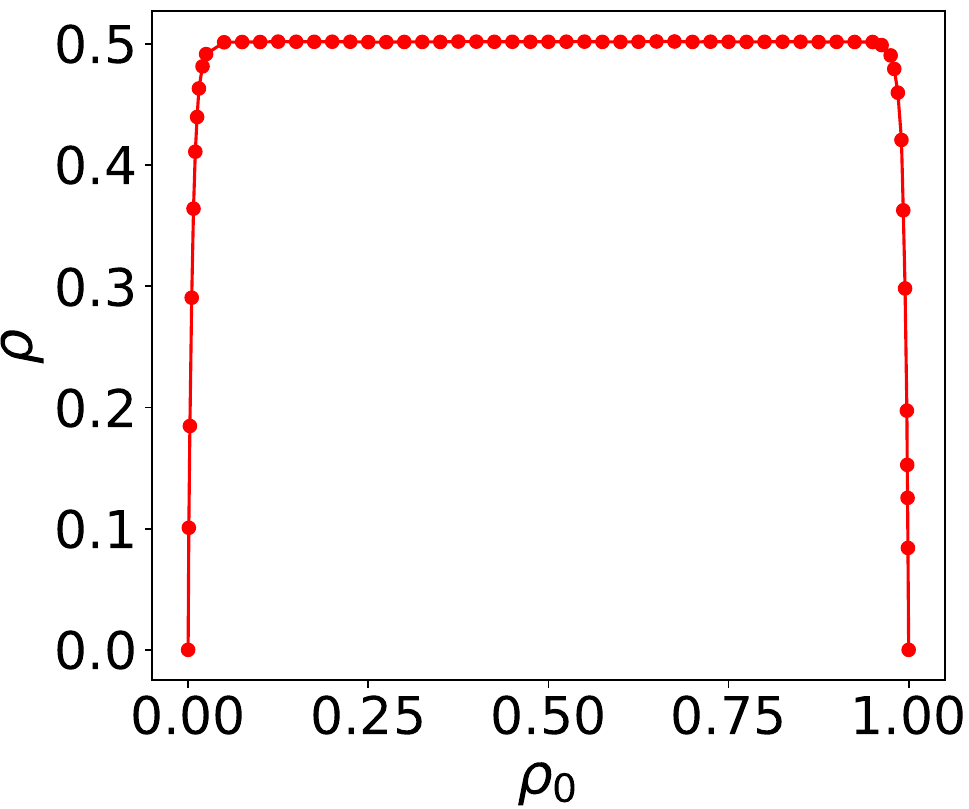}
\end{tabular}
\caption{Density $\rho$ of the rule 126 under long-time evolution for disordered initial states with density $\rho _0$. $L$ = 200, $t$ = 2000.}
\label{fig:Densities_Wolfram_rule_126}
\end{figure}

The rule 126, another complex rule, is simulated on arrays of size $L$ = 200 and $t$ = 2000, with the configurations from the last 200 time steps ($\Delta t$ = 200) being taken. For each initial density $\rho _0$, 1000 configurations are generated to calculate the density. As shown in Fig.~\ref{fig:Densities_Wolfram_rule_126}, the long-time density $\rho$ for disordered initial states is presented. Unlike the rules 18 and 90, the rule 126 exhibits a consistent asymptotic density $\rho_\infty =0.5 \pm 0.002$, across a broader range of initial densities. This result robustly holds under varying initial densities and system sizes. However, it contradicts the theoretical value of $\rho_\infty = 0.25$ derived in existing literature \cite{wolfram1983statistical,grassberger1983new}.

\begin{figure}[htbp]
\setlength{\tabcolsep}{0pt}
\centering
\begin{tabular}{cc}
\includegraphics[width=0.49\columnwidth]{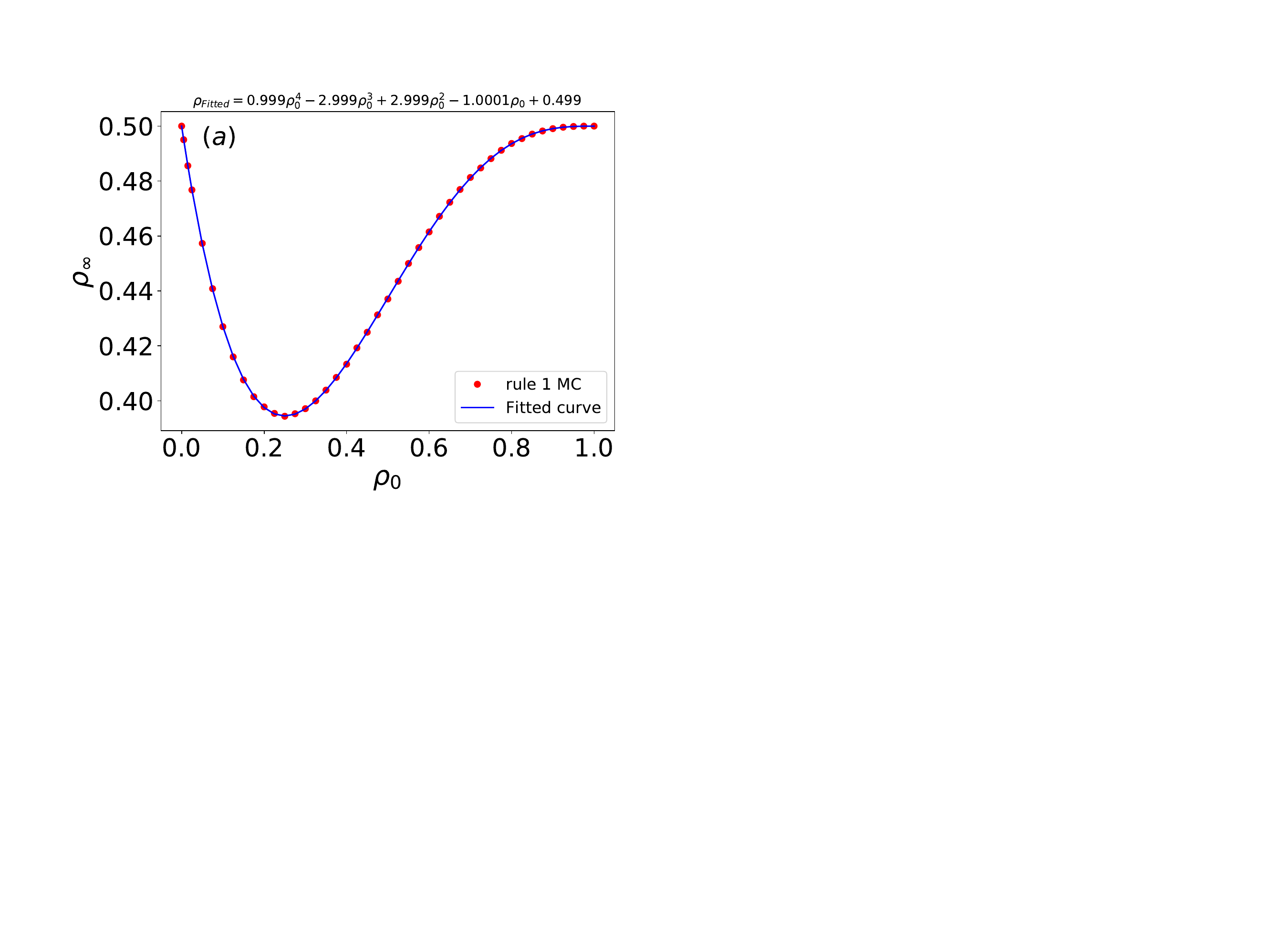} &
\includegraphics[width=0.49\columnwidth]{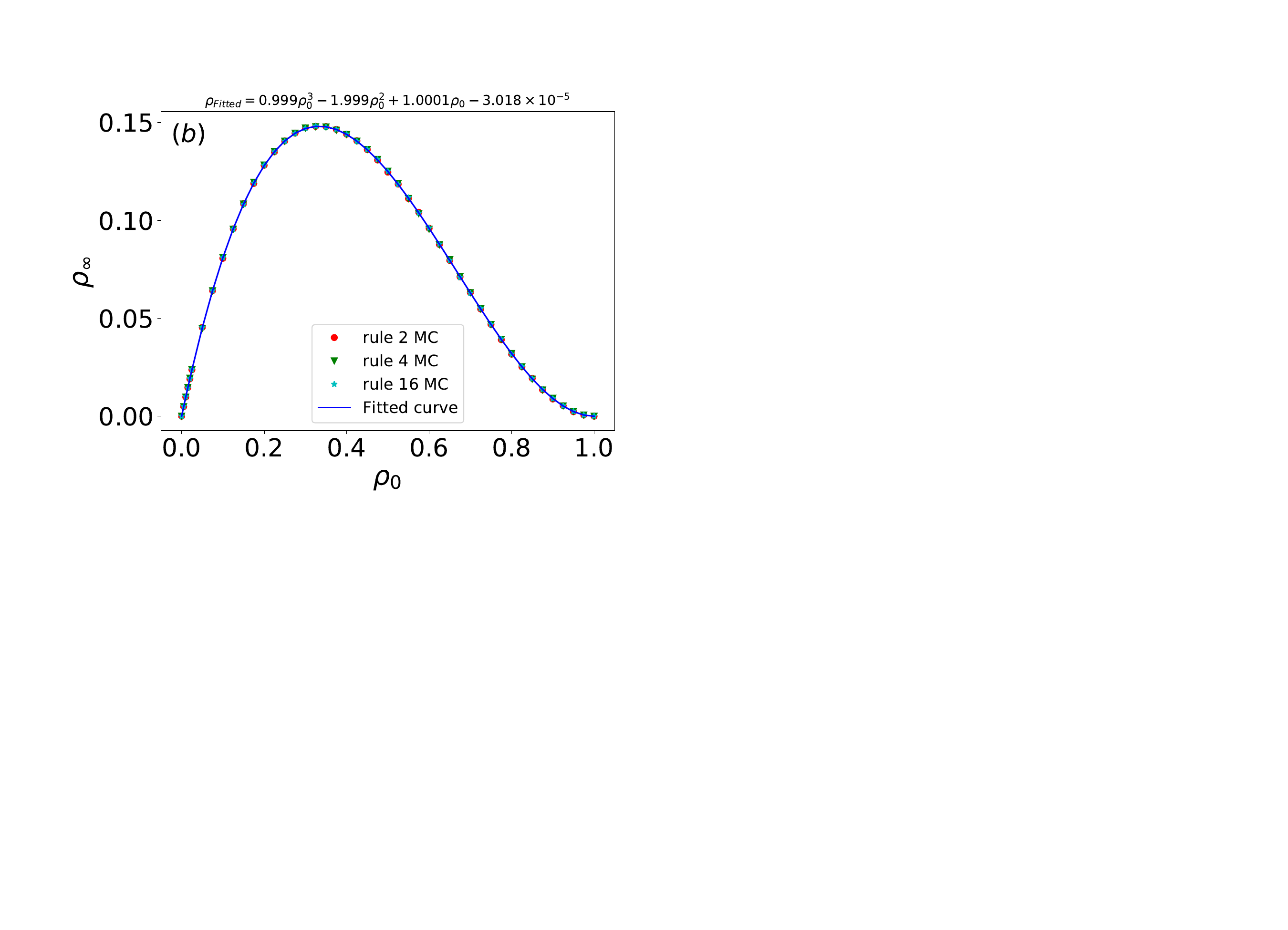} \\
\includegraphics[width=0.49\columnwidth]{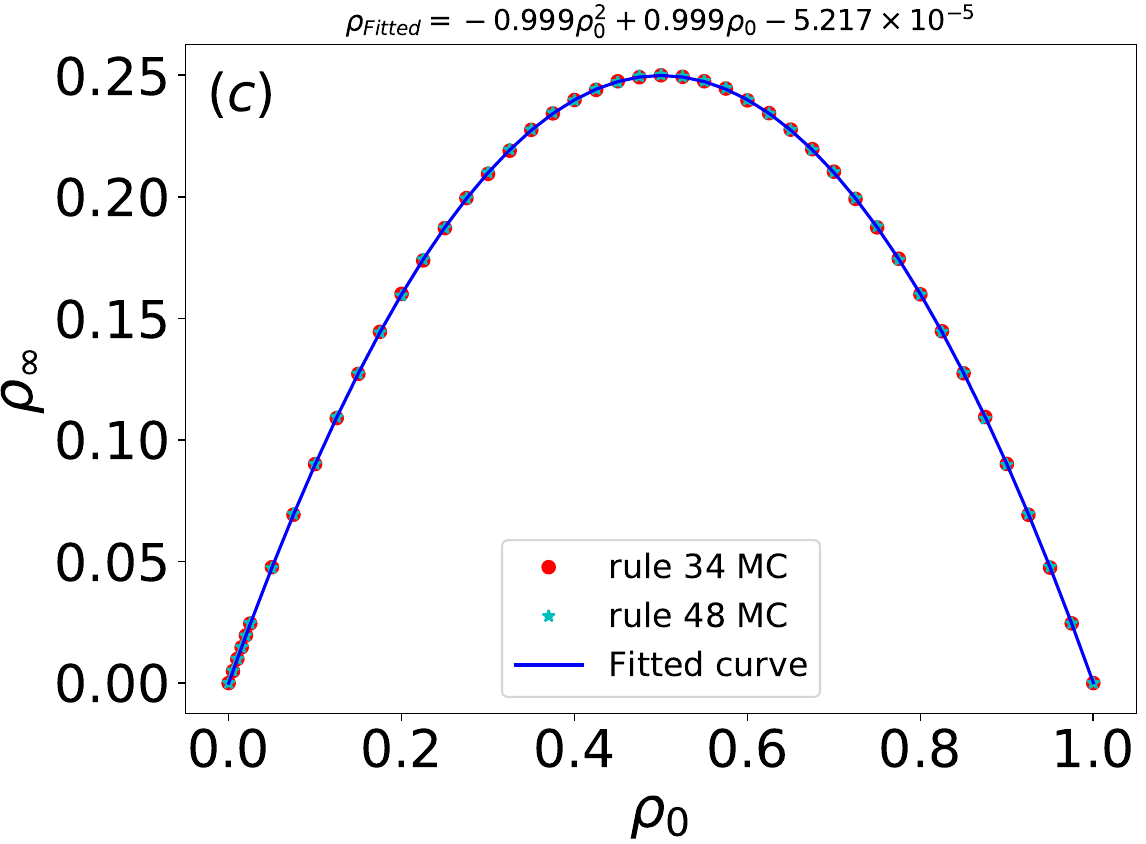}&
\includegraphics[width=0.49\columnwidth]{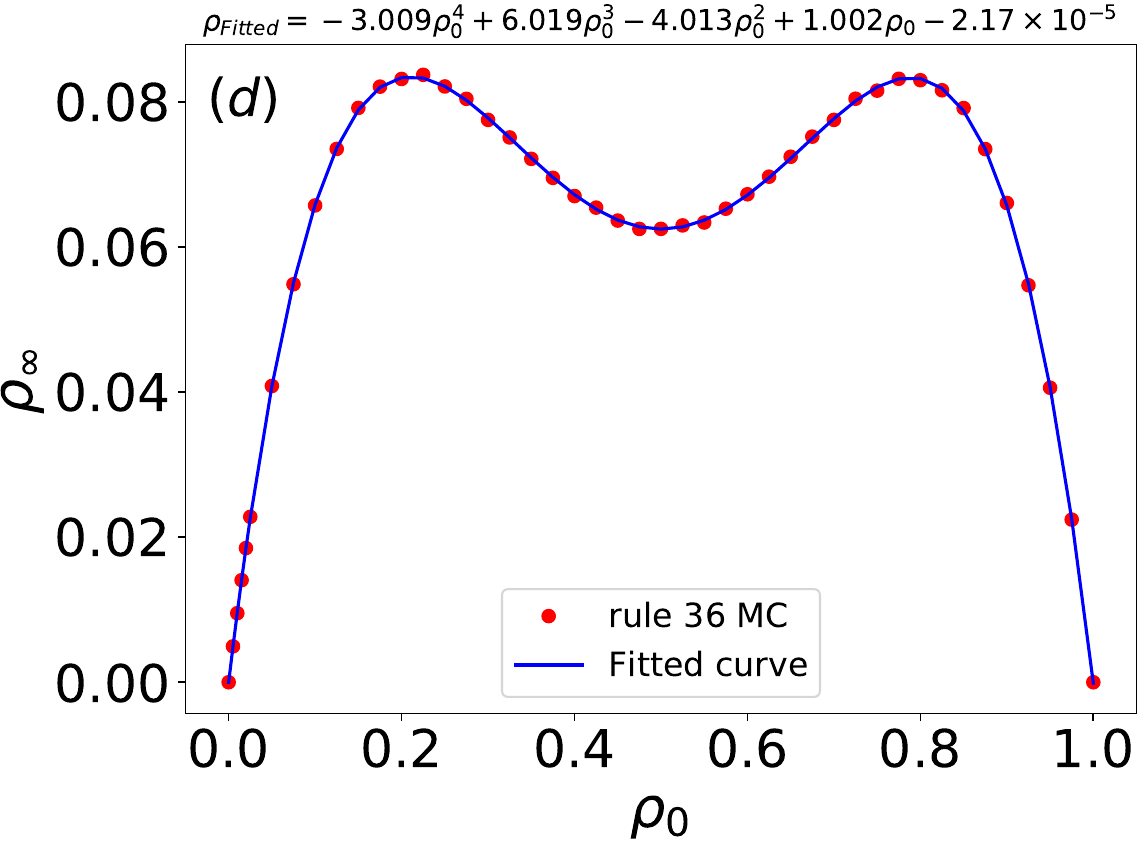}\\
\includegraphics[width=0.49\columnwidth]{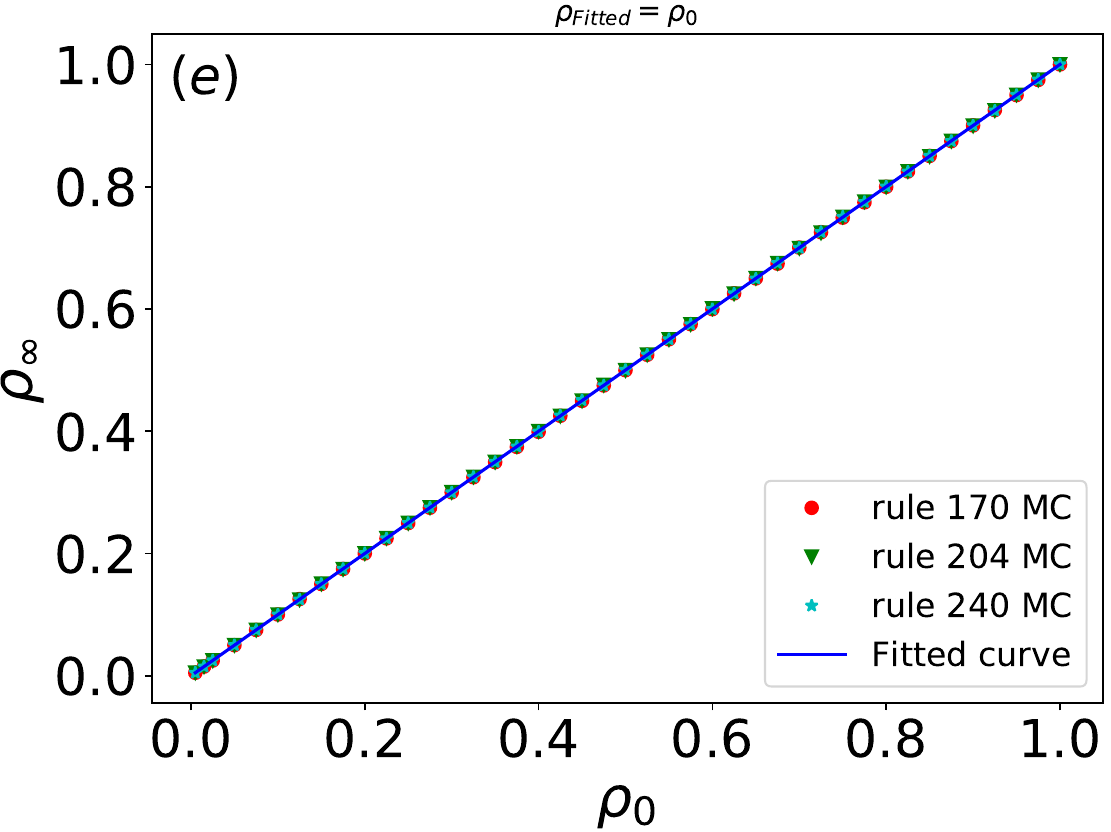} &
\end{tabular}
\caption{Asymptotic density $\rho_\infty$ of the rules 1, 2, 4, 16, 34, 48, 36, 170, 204, and 240 under disordered initial configurations with varying $\rho _0$, and fitting results for $\rho_\infty$. $L$ = 1000, $t$ = 30. Red dots represent asymptotic density of Wolfram rules and blue lines represent the fitting curve.}
\label{fig:rule_1_2_34_36_204_p_MC}
\end{figure}

\begin{table*}[htbp]
\centering
\caption{The Fitted density of the rules 1, 2, 4, 16, 34, 48, 36, 170, 204 and 240, and approximate asymptotic density $\rho_\infty$.}
\begin{tabular}{ccc}
\hline\hline
  & fitted density & approximate asymptotic density   \\
\hline
rule 1 & $\rho_{Fitted}=0.999\rho_0^4-2.999\rho_0^3 +2.999\rho_0^2-1.0001\rho_0+0.499$ &  $\rho_{\infty}$ = $\rho_0^4-3\rho_0^3+3\rho_0^2-p+0.5$  \\
\hline
rules 2, 4 and 16  &  $\rho_{Fitted} = 0.999\rho_0^3-1.999\rho_0^2 +1.0001\rho_0-3.018\times 10^{-5}$ & $\rho_\infty=\rho_0^3-2\rho_0^2+\rho_0$\\
\hline
rules 34 and 48  &  $\rho_{Fitted}=-0.999\rho_0^2 +0.999\rho_0-5.217\times 10^{-5}$ & $\rho_\infty=-\rho_0^2 +\rho_0$  \\
\hline
rule 36 & $\rho_{Fitted}=-3.01\rho_0^4+6.019\rho_0^3 -4.013\rho_0^2+1.002\rho_0-2.17\times 10^{-5}$ & $\rho_\infty=-3\rho_0^4+6\rho_0^3 -4\rho_0^2+\rho_0$ \\
\hline
rules 170, 204, and 240  & $\rho_{Fitted}=\rho_0$ &$ \rho_\infty=\rho_0$ \\
\hline\hline
\end{tabular}
\label{table:asymptotic density}
\end{table*}

For the rules 1, 2, 4, 16, 34, 48, 36, 170, 204, and 240, simulations are run on arrays of size $L$ = 1000, and time steps $t$ = 30, with the configurations from the last 10 time steps ($\Delta t$ = 10) being taken. For each initial density $\rho _0$, 1000 configurations are generated to calculate the density. The density $\rho$ of these simple Wolfram rules stabilizes rapidly, with the asymptotic density $\rho_\infty$ exhibiting dependence on the initial density $\rho _0$. As shown in Fig.~\ref{fig:rule_1_2_34_36_204_p_MC}, the asymptotic density $\rho_\infty$ 
 under disordered initial configurations with varying $\rho _0$ is presented for all rules, including parameter optimization and fitting results. A least squares fitting procedure yields fitted densities $\rho_{Fitted}$ for each rule, and approximate values of $\rho_\infty$ are summarized in Table.~\ref{table:asymptotic density}.
 
For the rule 1, whose evolution depends solely on the condition \texttt{000$\rightarrow$1}, the asymptotic density $\rho_\infty$ = 0.5 when $\rho _0$ is 0 or 1. Conventional analytical methods struggle to determine the extrema of such functions, but our approach employs Python's symbolic computation library (SymPy) to obtain precise solutions. Specifically, a minimum density of 0.395 occurs at $\rho _0 = 0.25$.

The asymptotic density curves for the rules 2, 4, and 16 coincide apart from minor fluctuations. A maximum density of 0.148 is at $\rho _0=0.333$. The evolution of the rule 2 depends only on \texttt{001$\rightarrow$1}, the rule 4 on \texttt{010$\rightarrow$1}, and the rule 16 on \texttt{100$\rightarrow$1}. Despite their distinct evolutionary paths under the same initial density, all three rules converge to the same asymptotic density, indicating functionally equivalent dynamical mechanisms. This suggests that these rules can serve as effective filters in certain applications.

Similarly, the rules 34 and 48 exhibit identical asymptotic density $\rho_\infty$. The rule 34 depends on condition \texttt{101$\rightarrow$1} and \texttt{001$\rightarrow$1}, the rule 48 depends on condition \texttt{101$\rightarrow$1} and \texttt{100$\rightarrow$1}. At $\rho _0=0.5$, both rules reach a maximum density of 0.25. 

For the rule 36, the fitted density curve displays two local maxima and one local minimum. The asymptotic density reaches a local minimum of 0.0625 at $\rho _0=0.5$, and local maxima of approximately 0.0833 occur at $\rho _0=0.211$ or $\rho _0=0.789$.

The rules 170, 204, and 240 all satisfy $\rho_\infty =\rho_0$. The rule 204 is the ``identity rule'', defined by the Boolean function $s_{i}(t+1) = s_i(t)$, meaning the configuration remains unchanged over time. The rule 170 is characterized by $s_{i}(t+1) = s_{i+1}(t)$, and the rule 240 by $s_{i}(t+1) = s_{i-1}(t)$. The asymptotic density of these rules always matches the initial density, confirming that they share identical dynamical evolution mechanisms.

The asymptotic densities of Wolfram automata rules have been extensively studied through numerical simulations. For complex rules such as the rules 18 and 90, the asymptotic density 
$\rho_\infty$ converges to a well-defined value under most disordered initial configurations with initial density $\rho _0$. However, under extreme initial conditions (e.g., $\rho _0 \leq 0.01$ or $\rho _0\geq 0.99$), even after extended temporal evolution (e.g., up to $t$ = $10^6$), the density $\rho_t$ fails to stabilize. Instead, it exhibits substantial fluctuations and significant deviations from the expected asymptotic density $\rho_\infty$.

In contrast, simpler Wolfram rules such as the rules 2, 4, 16, 34, and 48 stabilize rapidly within a small number of time steps, with the asymptotic density $\rho_\infty$ depending on the initial density $\rho _0$. Notably, despite being governed by distinct local update conditions, 
for instance, the rule 2 depends on \texttt{001$\rightarrow$1}, the rule 4 on \texttt{010$\rightarrow$1}, and the rule 16 on \texttt{100$\rightarrow$1}. These rules exhibit identical asymptotic densities under the same initial density $\rho _0$, even though their configurations evolve through different pathways over time. This convergence behavior suggests that they share equivalent underlying dynamical mechanisms, irrespective of differences in their local transition functions.

\section{Supervised learning of the Wolfram automata \label{sec:Wolfram Supervised}}

Supervised learning is applied to study Wolfram automata, with data comprising raw configurations generated from Monte Carlo (MC) simulations of the (1+1)-dimensional Wolfram automata. Within the (1+1)-dimensional Wolfram automata framework encompassing 256 distinct rules, certain rules rapidly converge to trivial steady states (e.g., homogeneous or empty configurations), enabling straightforward classification. We focus on Wolfram rules that exhibit complex structural configurations under disordered initial states during temporal evolution. Ten typical Wolfram rules are selected: rules 6, 16, 18, 22, 36, 48, 150, 90, 182, and 190. The generated configurations are divided into a training set and a test set, the configurations $x_i$ of each rule are labeled with an identical tag. For ten distinct Wolfram rules, the true label $\mathbf{y}_i$ is a \textbf{one-hot vector}. For instance, the rule 6 corresponds to \([1, 0, 0, \dots]\), and the rule 90 corresponds to \([\dots, 0, 1, 0, 0]\), where only the position indexing the specific rule is set to 1, and all others are 0.

\begin{figure}[htbp]
\setlength{\tabcolsep}{1.2pt}
\centering
\begin{tabular}{c}
\includegraphics[width=0.9\columnwidth]{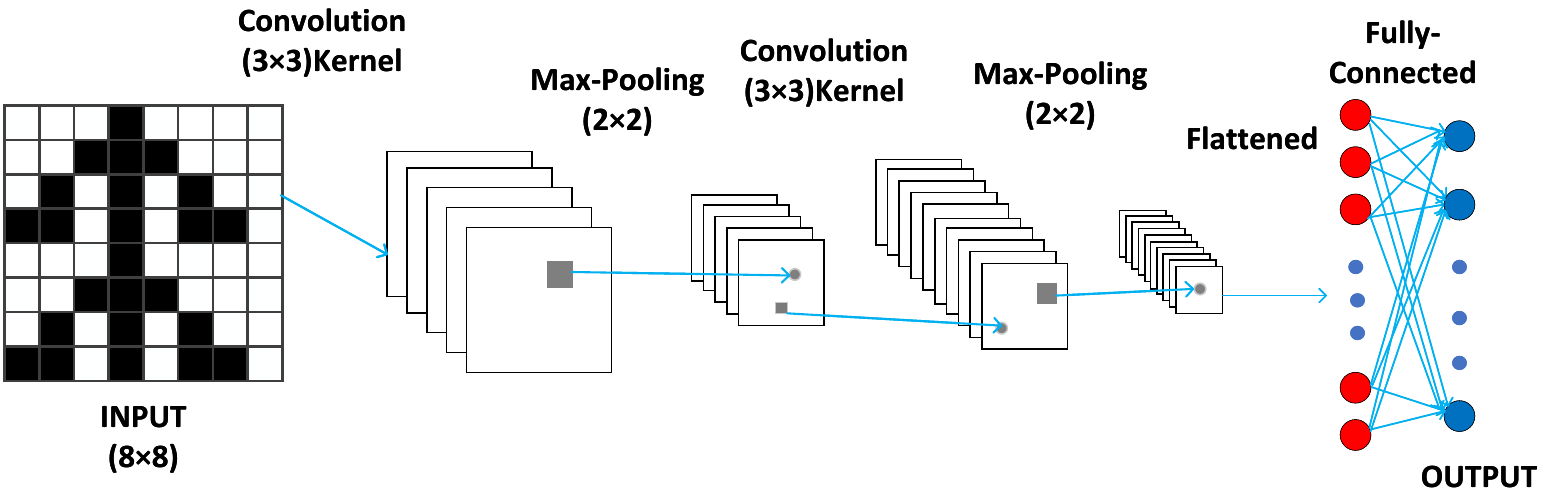}
\end{tabular}
\caption{The CNN architecture consists of two convolutional layers, each followed by a max-pooling layer, a fully connected layer, and a \textbf{softmax} output layer. The first convolutional layer applies 3$\times$ 3 \textbf{kernels} with \textbf{sigmoid} activation to extract spatial features, followed by 2$\times$2 max-pooling for dimensionality reduction. The second convolutional layer repeats this pattern, further refining feature maps. The pooled features are then flattened and passed through a fully connected layer using \textbf{sigmoid} activation. Finally, the output layer employs \textbf{softmax} activation to produce class probabilities for classification tasks. }
\label{fig:Wolfram Schematic CNN}
\end{figure}

For supervised learning, we apply the convolutional neural network (CNN) as illustrated in Fig.~\ref{fig:Wolfram Schematic CNN}. The CNN architecture consists of two convolutional layers, each followed by a max-pooling layer, a fully connected layer, and a \textbf{softmax} output layer. The first convolutional layer applies 3$\times$ 3 \textbf{kernels} with \textbf{sigmoid} activation, extracting spatial features while introducing nonlinearity, followed by 2$\times$2 max-pooling for dimensionality reduction, enhancing translational invariance. The second convolutional layer repeats this pattern, further refining feature maps. The pooled features are then flattened and passed through a fully connected layer using \textbf{sigmoid} activation. Finally, the output layer employs \textbf{softmax} activation to produce class probabilities for classification tasks. 

Since Learning machines extract features from configuration images generated across distinct rules and small system sizes, selecting configurations that maximize information capture is essential. For complex Wolfram automata such as the rules 18 and 90, under low or high initial densities (e.g., $\rho _0 \leq 0.01$ or $\rho _0\geq 0.99$), the density evolving over time remains low; thus, it is customary to select random initial states (probability of 1 per site is 0.5) for simulations to capture non-trivial dynamical behaviors. For simple rules, the asymptotic density $\rho_\infty$ depends on the initial density. Therefore, the initial density $\rho_0$ corresponding to a high asymptotic density should be optimally selected. For example, the rule 16 adopts an initial state density $\rho_0$ of 0.33, while the rule 36 sets it to 0.21.

The configuration images are of $L\times (t + 1)$ dimension, $L=30$, $t=100$. For each rule, 2000 labeled configurations are generated for the training set and another 200 configurations for the test set. The CNN output layers are eventually averaged over the test set.

Categorical cross-entropy loss is the standard choice for data classification tasks in machine learning. For Wolfram rule classification, it thus serves as the designated loss function. The Categorical cross-entropy loss function is defined as:
\begin{equation}
\mathcal{L} = -\frac{1}{N} \sum_{i=1}^{N} \log(\hat{y}_{i, k_i}),
\end{equation}
$N$ is the number of samples in the batch (batch size), $\hat{y}_{i, k_i}$ is the model's predicted probability for the true class label $k_i$ of sample $i$.

\begin{figure}[htbp]
\setlength{\tabcolsep}{1.2pt}
\centering
\begin{tabular}{cc}
\includegraphics[width=0.493\columnwidth]{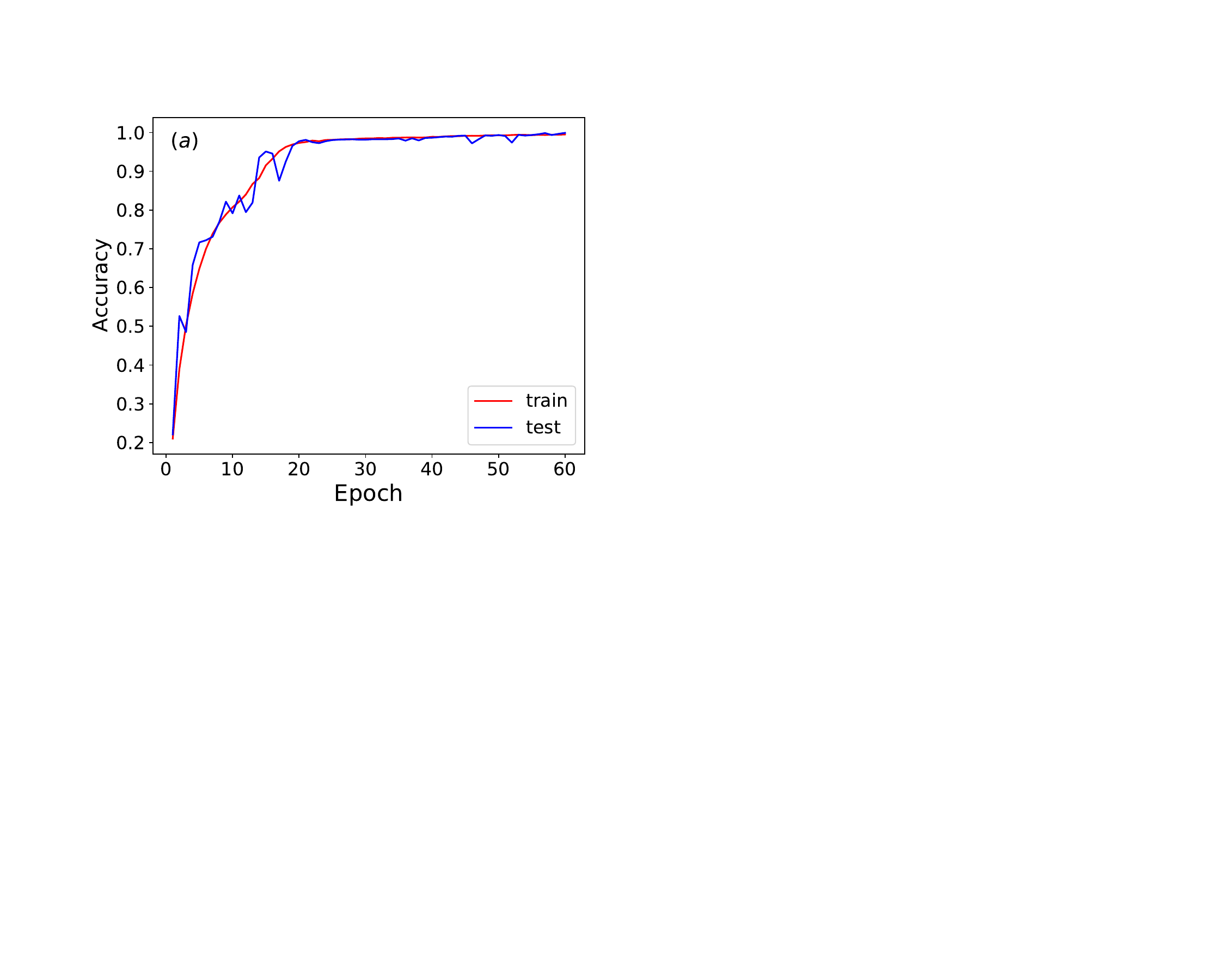}&
\includegraphics[width=0.481\columnwidth]{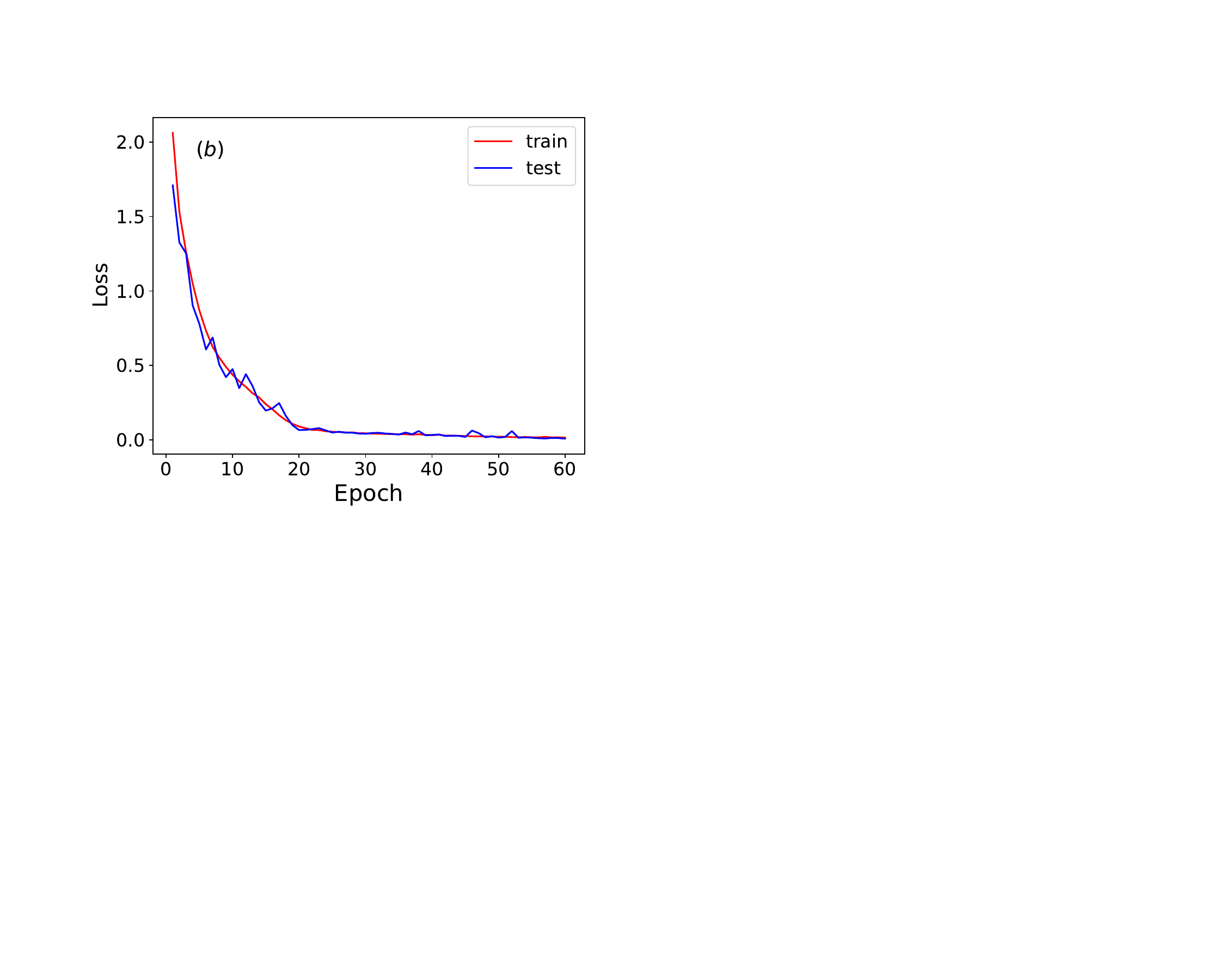}\\
\end{tabular}
\caption{The accuracy and loss function of the trained neural network across epochs. }
\label{fig:Wolfram_rule_Supervised_Acc_loss}
\end{figure}

As illustrated in Fig.~\ref{fig:Wolfram_rule_Supervised_Acc_loss}, the accuracy and loss function of the trained neural network stabilize around epoch 60. The CNN achieves an impressive $99.9\%$ test accuracy in classifying Wolfram rules. This high level of recognition accuracy highlights the CNN's exceptional ability to reliably classify Wolfram rules, even within complex systems.

\begin{figure}[htbp]
\setlength{\tabcolsep}{1.2pt}
\centering
\begin{tabular}{cc}
\includegraphics[width=0.45\columnwidth]{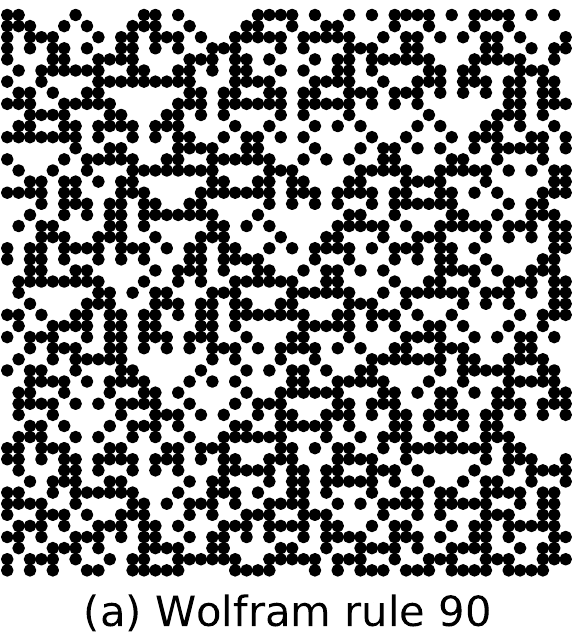}&
\includegraphics[width=0.441\columnwidth]{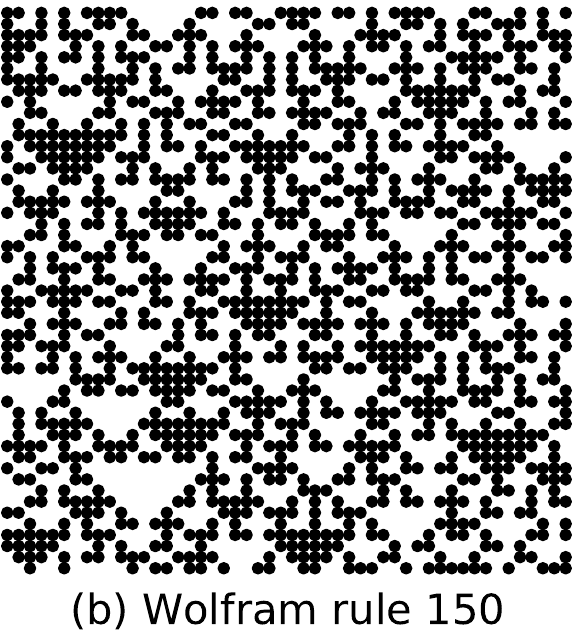}\\
\end{tabular}
\caption{Configurations of the rules 90 and 150 under random initial conditions. }
\label{fig:Wolfram_rule_90_150_random}
\end{figure}

As shown in Fig.~\ref{fig:Wolfram_rule_90_150_random}, under random initial states, the configurations of the rules 90 and 150 exhibit identical asymptotic density of 0.5. Over time, their evolved configurations become highly similar, making manual distinction challenging. However, a trained CNN achieves high classification accuracy in reliably identifying these rules. Although supervised learning provides prior knowledge of Wolfram rules, the neural network model can rapidly distinguish large volumes of Wolfram rule configurations with minimal training time—demonstrating a key advantage of machine learning for complex spatial-temporal pattern recognition tasks.

We employ a CNN as a supervised learning framework, due to its high effectiveness in image-like classification tasks, which offers superior efficiency and accuracy over traditional methods. By utilizing only a two-layer CNN architecture rather than deeper networks, we achieve optimal performance for Wolfram automata configurations with small system dimensions. This design minimizes trainable parameters, reduces training time, and maintains high classification accuracy.

\section{Unsupervised learning of the Wolfram automata \label{sec:Wolfram Unsupervised}}

Unsupervised learning operates on unlabeled data, enabling autonomous discovery of latent structures and classifications without human guidance—particularly valuable when labeled data is limited or costly to acquire. This section will apply two prominent unsupervised techniques, namely autoencoders and principal component analysis (PCA), to study Wolfram automata. Outputs are dimensionally reduced to two- and one-dimensional spaces, prioritizing maximal physical relevance, a conclusion directly supported by evidence in this section.

\subsection{Autoencoder results of the Wolfram automata}

\begin{figure}[htbp]
\setlength{\tabcolsep}{1.2pt}
\centering
\begin{tabular}{c}
\includegraphics[width=0.7\columnwidth]{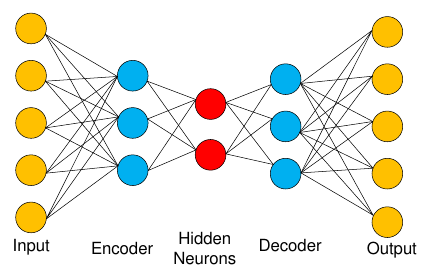}
\end{tabular}
\caption{Schematic diagram of the autoencoder architecture. The fully connected autoencoder structure comprises an input layer, an encoder with four neural layers, a latent layer containing one or two neurons, a decoder with five neural layers, and an output layer. The neuron counts per layer follow the sequence: (256, 128, 64, 16, 2/1, 16, 64, 128, 256, 640), with ReLU activation functions applied throughout. Mean Squared Error (MSE) is selected as the loss function, and stochastic gradient descent (SGD) as the optimizer.}
\label{fig:Autoencoder automata}
\end{figure}

Autoencoders learn compressed representations of input data to reconstruct inputs and generate new samples resembling the original data distribution \cite{liou2014autoencoder, pinheiro2021variational, pinaya2020autoencoders}. As depicted in Fig.~\ref{fig:Autoencoder automata}, the fully-connected autoencoder architecture implemented in our study comprises an input layer, an encoder, a latent layer containing hidden neurons, a decoder, and an output layer. The inputs for the autoencoders are just raw configurations of
Wolfram automata. The Autoencoder employs the Mean Squared Error (MSE) as its loss function, its core objective is to minimize the discrepancy between input data and reconstructed data, thereby learning an effective low-dimensional representation (latent representation) of the data. Through this compression-decompression mechanism, the latent layer generates a compressed representation that retains essential input features, enabling the decoder to reconstruct the original data patterns by reversing the dimensionality reduction process.

First, select four Wolfram rules: 18, 90, 182, and 190. For the Wolfram rules, simulations are run on arrays of size $L=30$ and $t=150$, with the configurations from the last 50 time steps ($\Delta t =50$) being taken. For each rule, 2000 configurations are generated for the training set, and another 1000 configurations for the test set. By constraining the autoencoder's output to two neurons, the configurations of Wolfram automata are compressed into two dimensions. After training is completed, each input $\mathbf{x}_i$ from the test set produces a point $(h_{i1},h_{i2})$ on the two-dimensional plane.

\begin{figure}[htbp]
\setlength{\tabcolsep}{1.2pt}
\centering
\begin{tabular}{c}
\includegraphics[width=0.7\columnwidth]{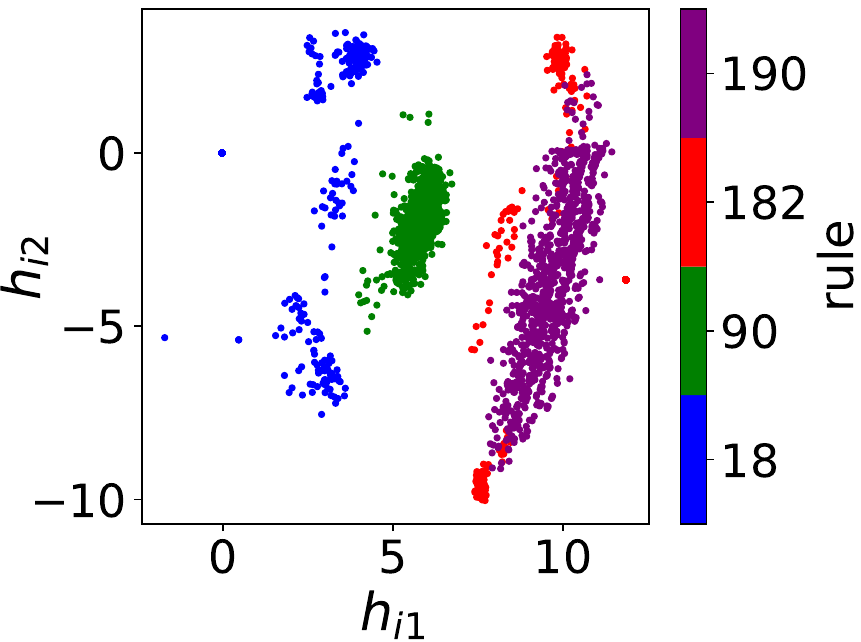}
\end{tabular}
\caption{The autoencoder's encoder output, featuring two hidden layers, compresses configurations generated from the rules 18, 90, 182, and 190 into a two-dimensional latent space. A color map distinguishes different Wolfram rules in the visualization.}
\label{fig:Wolfram_18_90_182_190_Autocode_1}
\end{figure}

As illustrated in the Fig.~\ref{fig:Wolfram_18_90_182_190_Autocode_1}, the autoencoder compresses configurations of Wolfram automata into a two-dimensional latent space. Apart from partial overlap in configurations of the rules 182 and 192, the autoencoder effectively separates configurations of distinct rules. Moreover, configurations of identical rules are clustered together, demonstrating the autoencoder’s capability to distinguish different Wolfram rules based on their spatiotemporal characteristics.

We also examine the rules 34 and 90, with configuration data sampled under identical conditions to those used for the aforementioned rules. By constraining the hidden layer to a single neuron, we investigate whether the features learned by the autoencoder correspond to the density of the Wolfram automata.

\begin{figure}[htbp]
\setlength{\tabcolsep}{1.2pt}
\centering
\begin{tabular}{cc}
\includegraphics[width=0.49\columnwidth]{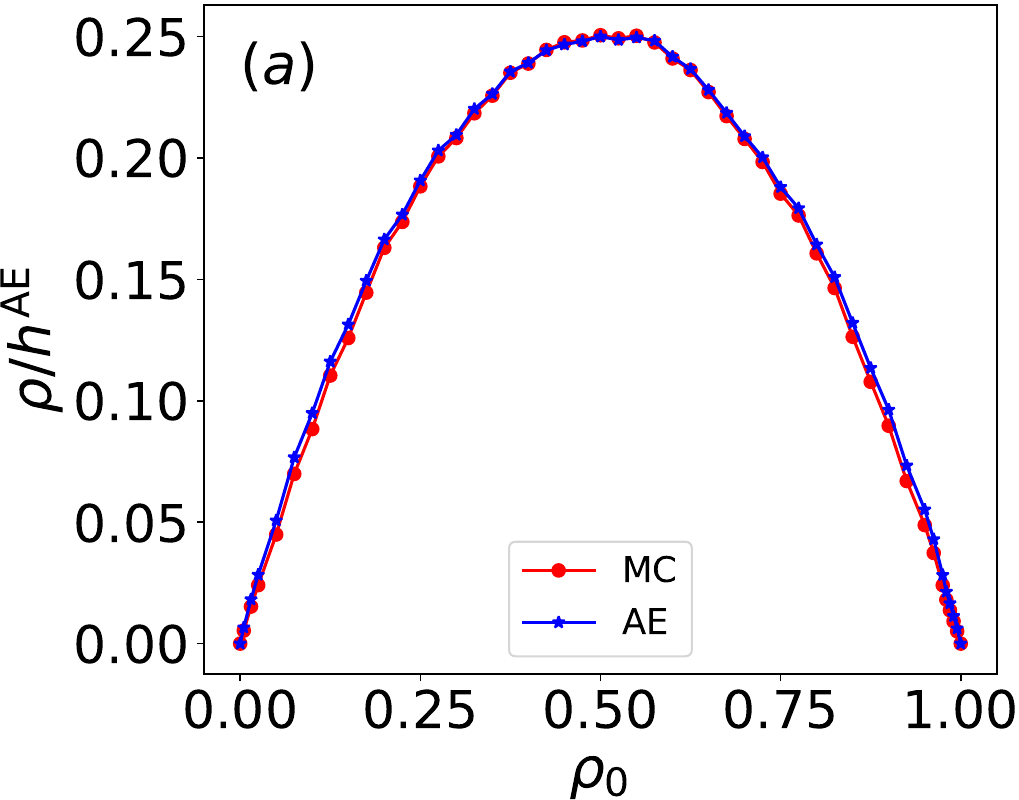}&
\includegraphics[width=0.49\columnwidth]{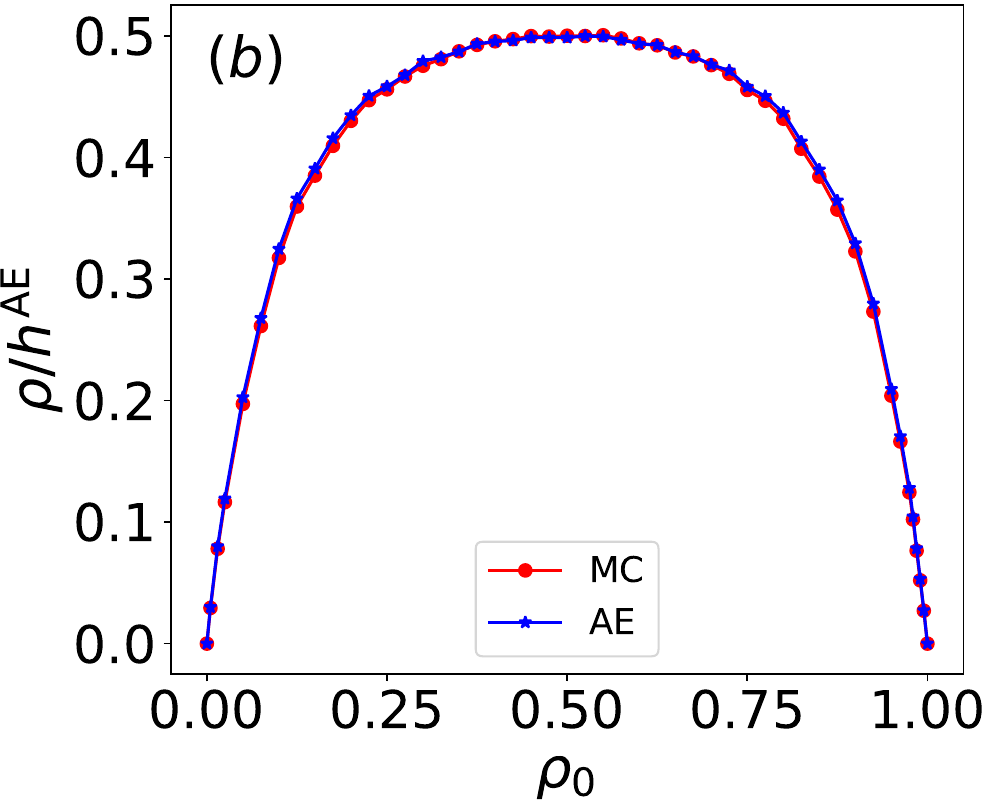}\\
\end{tabular}
\caption{The raw configurations of Wolfram automata  are encoded on to a latent variable of a single hidden neuron as a function of the initial density $\rho _0$ for (a) the rule 34, (b) the rule 90. The single-variable outputs (blue) of Wolfram automata autoencoder results are normalized, which are compared to the density from the Monte Carlo simulations. Normalized latent outputs (blue) exhibit quantitative agreement with the density (red) derived from Monte Carlo simulations.}
\label{fig:Autoencoder_rule_34_90_SGD}
\end{figure}

As shown in the Fig.~\ref{fig:Autoencoder_rule_34_90_SGD}, the regularized output $h^\mathrm{AE}$ for the rules 34 and 90 varies as a function of the initial density $\rho _0$. For the rule 34, the regularized output $h^\mathrm{AE}$ closely aligns with the density except at a few isolated points. In contrast, for the rule 90, $h^\mathrm{AE}$ exhibits stronger agreement with the density, indicating that the regularized single latent variable can be interpreted as the density. The observed discrepancy for the rule 34 arises because its density is a quadratic function of the initial density (i.e., nonlinear and complex relationship) rather than a simple linear dependence. The autoencoder fails to fully capture this nonlinear characteristic, leading to deviations between $h^\mathrm{AE}$ and the density at specific values of $\rho _0$ .

\subsection{PCA results of the Wolfram automata}

Principal Component Analysis (PCA) \cite{mackiewicz1993principal, greenacre2022principal, bharadiya2023tutorial}, as an unsupervised learning algorithm, reduces data dimensionality through orthogonal transformation. It identifies orthogonal directions of maximum variance in the data space, converting potentially correlated variables into linearly uncorrelated principal components (PCs). For Wolfram automata analysis, we focus exclusively on the first two principal components, which capture the dominant sources of variation.

This process can be conceptualized as projecting high-dimensional data onto a lower-dimensional manifold—specifically, selecting projection axes that maximize retained variance. Such projections preserve maximal information integrity after dimensionality reduction. In alignment with prior methodologies, we prioritize reductions to a physically interpretable low-dimensional space.

The PCA methodology is implemented through the following specific procedure. Simulations are run on arrays of size $L=30$ with temporal evolution spanning $t=150$ steps. From each simulation, configurations are extracted exclusively from the last $50$ time steps ($\Delta t = 50$). These configurations are structured into $M$-dimensional vectors $\mathbf{x}_i$, where $M = L \times \Delta t$. For each distinct rule, $1000$ such configurations are generated, resulting in a comprehensive dataset of $N = 1000$ samples. The full dataset is organized into an $(N \times M)$-dimensional matrix $\mathbf{X} = (\mathbf{x}_1, \mathbf{x}_2, \dots, \mathbf{x}_i, \dots, \mathbf{x}_{N})^T$, which serves as the foundational input for PCA computations.  

PCA operates by determining principal components through a linear transformation $\mathbf{Y} = \mathbf{X}\mathbf{W}$. Here, the transformation matrix $\mathbf{W} = (\mathbf{w}_1, \mathbf{w}_2, \dots, \mathbf{w}_K)$ has dimensions $(M \times K)$, with each column vector $\mathbf{w}_n$ representing a weighted component of dimension $M$. When $K \ll M$, this transformation achieves significant dimensionality reduction.  

The principal component (PC) directions are derived by analyzing the real symmetric covariance matrix $\mathbf{X}^T\mathbf{X}$, which has dimensions $(M \times M)$. For the case $K = M$, the directions $\mathbf{w}_n$ correspond precisely to the eigenvectors of $\mathbf{X}^T\mathbf{X}$, satisfying the eigenvalue equation:  
\begin{equation}  
\mathbf{X}^T\mathbf{X}\mathbf{w}_n = \lambda_n \mathbf{w}_n \, . 
\label{equ: SVD}  
\end{equation}  
The eigenvalues $\lambda_n$ are sorted in descending order ($\lambda_1 \geq \lambda_2 \geq \dots \geq \lambda_M \geq 0$), quantifying the variance of $\mathbf{X}$ along each eigenvector direction. Dimensionality reduction is subsequently achieved by retaining only the first few eigenvectors $\mathbf{w}_n$ associated with the largest $\lambda_n$. In standard PCA terminology, the normalized eigenvalues $\tilde{\lambda}_n = \lambda_n / \sum_{i=1}^{M} \lambda_i$ define the \textbf{explained variance ratio}. This ratio represents the proportion of total variance captured by the $n$-th principal component, thereby quantifying its statistical significance in the reduced-dimensional representation.  

\begin{figure}[htbp]
\setlength{\tabcolsep}{1.2pt}
\centering
\begin{tabular}{c}
\includegraphics[width=0.7\columnwidth]{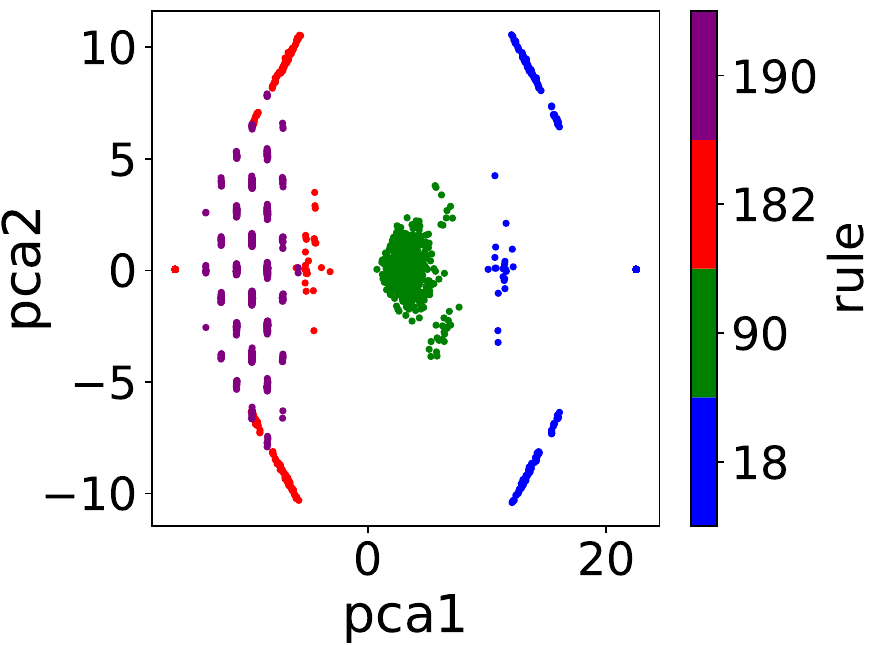}
\end{tabular}
\caption{PCA retains two principal components to compress the configurations of the rules 18, 90, 182, and 190 into two dimensions. A color map distinguishes different Wolfram rules in the visualization.}
\label{fig:Wolfram_18_90_182_190_PCA_1}
\end{figure}

For the rules 18, 90, 182, and 190, PCA compresses the configurations of Wolfram automata into two dimensions by retaining two principal components, as illustrated in the Fig.~\ref{fig:Wolfram_18_90_182_190_PCA_1}. Similar to autoencoder results, PCA effectively separates configurations of different Wolfram rules into distinct clusters. Additionally, configurations belonging to the same rule exhibit symmetric patterns within their clusters, demonstrating PCA’s capability to distinguish among various Wolfram rules based on their structural features

\begin{figure}[htbp]
\setlength{\tabcolsep}{1.2pt}
\centering
\begin{tabular}{cc}
\includegraphics[width=0.49\columnwidth]{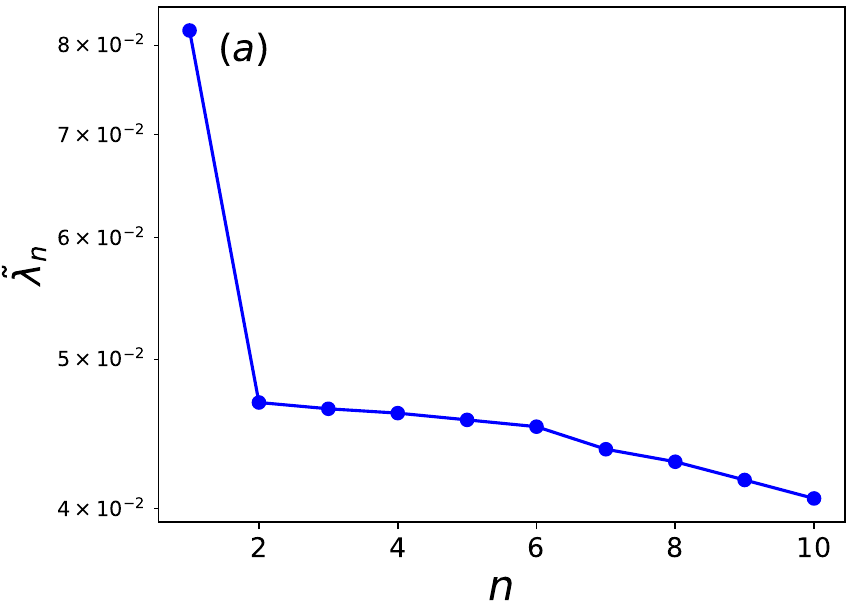}&
\includegraphics[width=0.481\columnwidth]{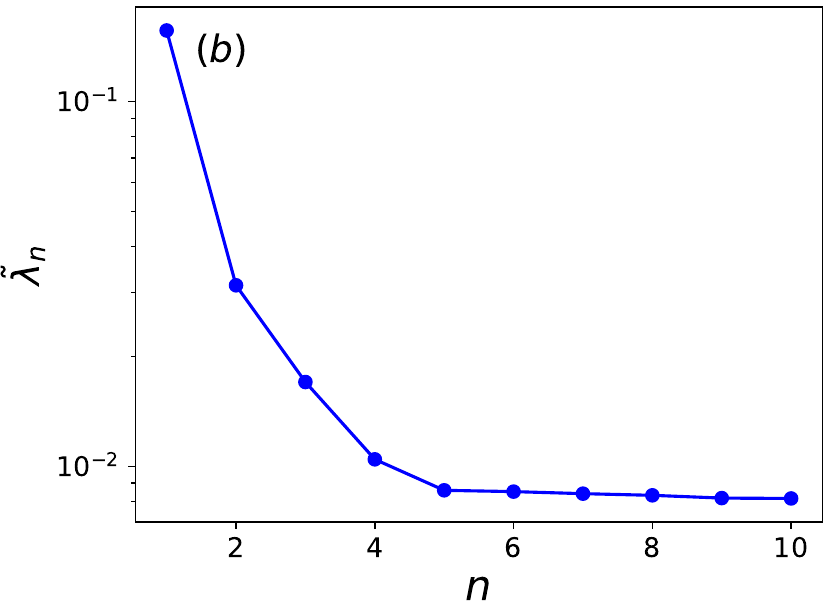}\\
\includegraphics[width=0.49\columnwidth]{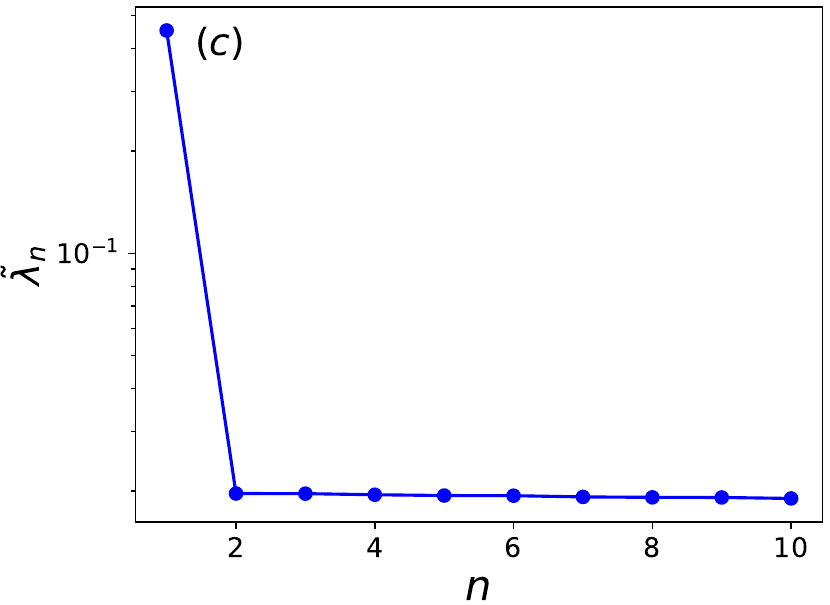}&
\includegraphics[width=0.49\columnwidth]{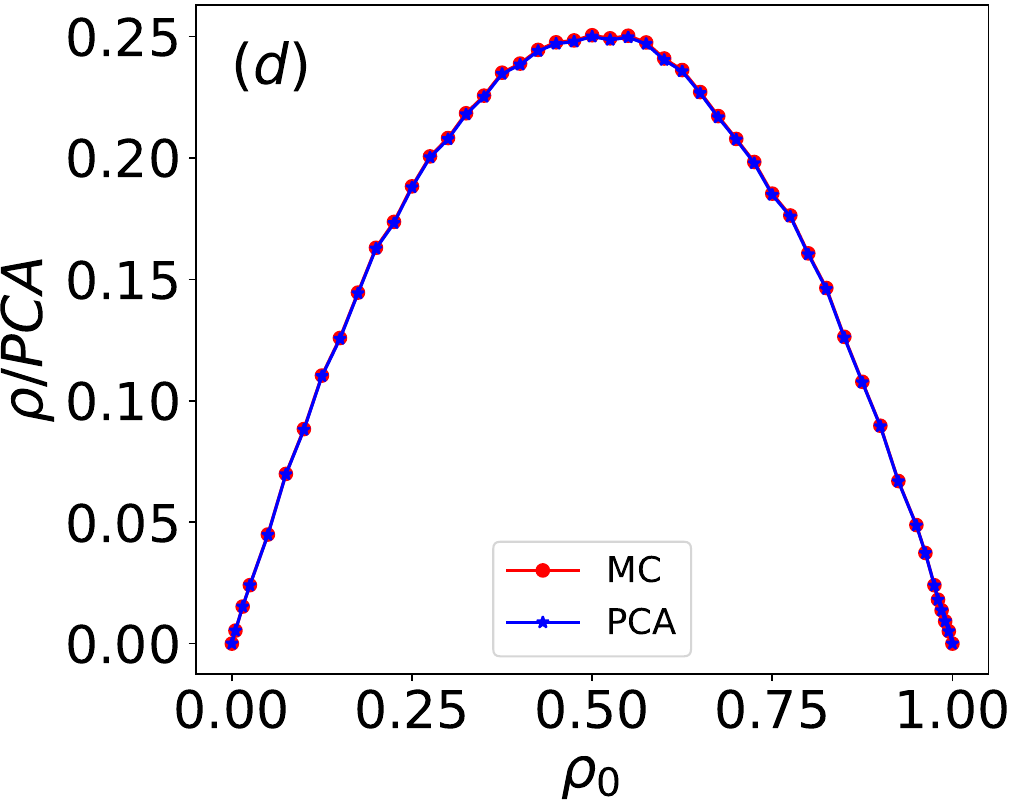}\\
\includegraphics[width=0.49\columnwidth]{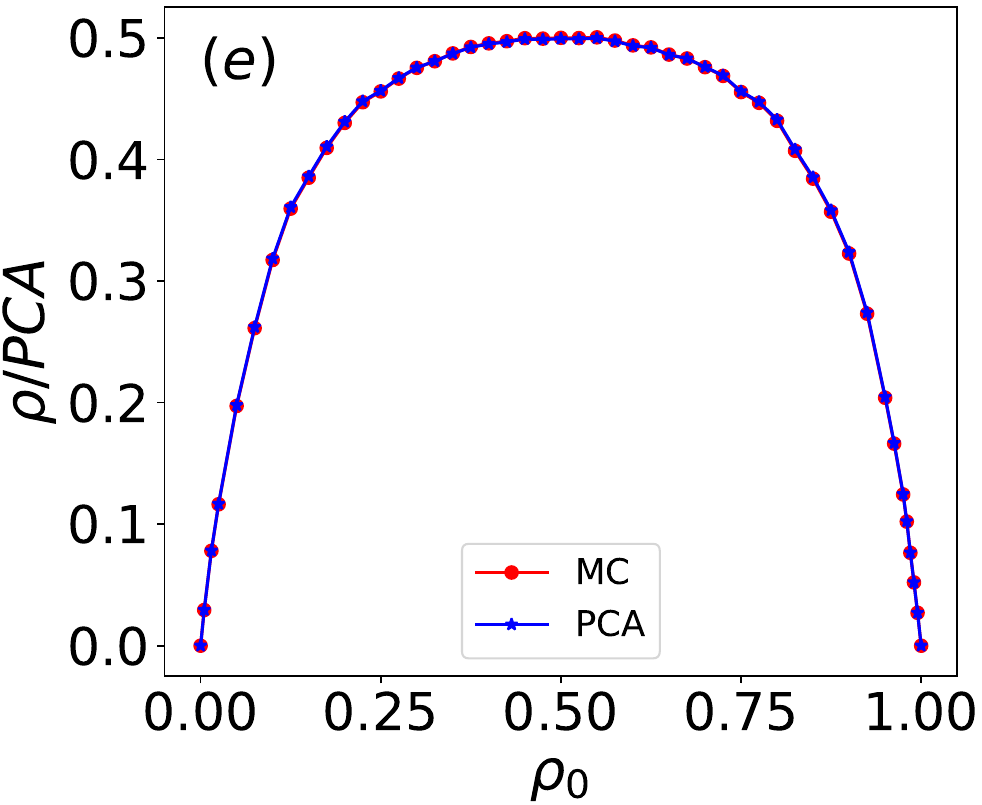}&
\includegraphics[width=0.49\columnwidth]{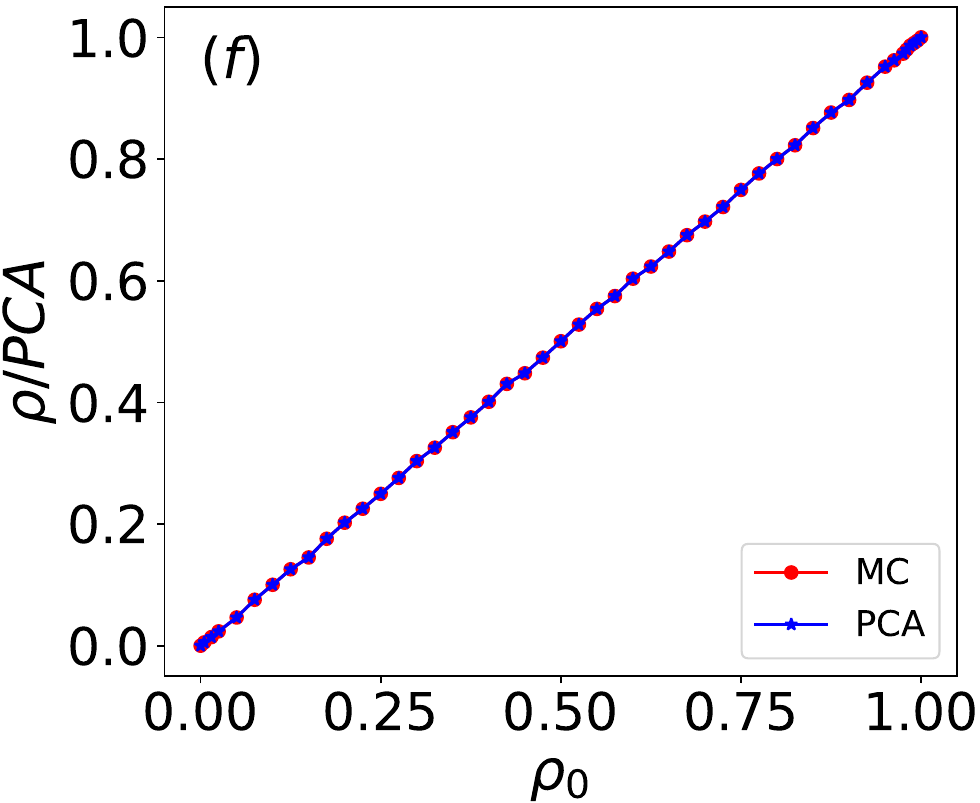}\\
\end{tabular}
\caption{PCA explained variance ratios of the first 10 principal components for (a) the rule 34, (b) the rule 90, and (c) the rule 204. PCA results for (d) the rule 34, (e) the rule 90 and (f) the rule 204. The normalized first principal component (blue) varies as a function of the initial density $\rho _0$, which is compared to the density (red) from the Monte Carlo simulations.}
\label{fig:variance_ratio_PCA_1_rule_34_90_204}
\end{figure}

As shown in Figs.~\ref{fig:variance_ratio_PCA_1_rule_34_90_204} (a) and (b), the variance ratios of the first and second principal components (PC1 and PC2) differ significantly across rules. For the rule 34, the variance ratios are 0.0818 (PC1) and 0.04 (PC2), while for the rule 90, they are 0.156 (PC1) and 0.0313 (PC2). In both cases, PC1 contributes the most to the total variance, yet its ratio remains relatively small. The variance explained ratio of the PC1 and PC2 is generally higher for linear data, essentially because the linear nature of PCA aligns well with the inherent structure of linear data.  The density $\rho$ of the rule 34 is a quadratic function of the initial density $\rho _0$, indicating strong nonlinearity in its dynamics. For nonlinear data, however, the linear assumption of PCA limits its ability to capture major variations, resulting in a more dispersed and lower variance explained ratio. This explains the suppressed variance ratio of PC1 (0.0818) and the similar ratio (0.04) for subsequent components in the rule 34. For the rule 90, although $\rho$ also exhibits nonlinear dependence on $\rho _0$, this relationship is weaker. Thus, PCA more effectively concentrates variance into PC1, yielding a higher ratio (0.156). The residual nonlinearity distributes into minor PCs (e.g., PC2 ratio = 0.0313), consistent with PCA’s linearity constraints. 

In contrast, as shown in Figure \ref{fig:variance_ratio_PCA_1_rule_34_90_204} (c), the variance ratio of PC1 for the rule 204 is 0.45, which is substantially higher than those of Rules 34 and 90. Since the density of the rule 204 maintains a linear relationship with the initial density, the variance is more effectively captured by the first principal component.

Furthermore, Figs.~\ref{fig:variance_ratio_PCA_1_rule_34_90_204} (d), (e), and (f) compare the normalized first principal component with the density for the rules 34, 90, and 204. The normalized first principal component of PCA closely matches the density, with minimal discrepancy observed across all cases, indicating that the regularized first principal component can be treated as the density for Wolfram automata.

The regularized single latent variable $h^\mathrm{AE}$ in autoencoders can approximate the density, while for certain Wolfram rules, deviations exist between the regularized output $h^\mathrm{AE}$ and the initial density $\rho _0$. In contrast, the normalized first principal component in PCA exhibits higher consistency with the density. Concurrently, PCA achieves shorter computation time and greater efficiency.

\section{\protect Summary}
\label{sec:Summary}
In this paper, we applied numerical simulations, computation methods, and supervised and unsupervised learning methods to study the asymptotic density and dynamical evolution mechanisms of the Wolfram automata. Through numerical simulations, we extend beyond single-site initial states to specific sequences, including rule 146 with ``\dots 0, 1, 1, 0, 1, 1, 0, 1, 1, 1, 0, 1, 1, 0, 1, 1 \dots'' and rule 154 with ``\dots 0, 1, 0, 1, 0, 1, 1, 0, 1, 0, 1 \dots''. the rules 26, 82, 146, 154, and 210 also evolve into Sierpinski triangle patterns over time.

The asymptotic density of Wolfram rules was numerically determined. For complex rules such as rules 18 and 90, most disordered initial configurations with densities $\rho_0$ densities $\rho_\infty$. However, under extremely low or high initial densities (e.g., $\rho _0 \leq 0.01$ or $\rho _0\geq 0.99$), even after extended evolution (e.g.,$t$=$10^6$), $\rho_t$ fails to stabilize. Instead, it exhibits substantial fluctuations and significant deviation from the expected asymptotic density. The density $\rho$ of simple Wolfram rules stabilizes within a small number of time steps, with the asymptotic density $\rho_\infty$ depending on the initial density $\rho _0$. Certain rules, such as rules 2, 4, and 16, or rules 34 and 48, rely on different local update conditions. Yet, Under identical initial density, even though their configurations evolve through different pathways over time, they converge to the same asymptotic density, suggesting that they share equivalent underlying dynamical mechanisms.

With supervised learning, a trained CNN achieves high accuracy in identifying these rules. Although supervised learning provides prior knowledge of Wolfram rules, the neural network model can rapidly distinguish large volumes of Wolfram rule configurations with minimal training time—demonstrating a key advantage of machine learning for complex spatial-temporal pattern recognition
tasks.

The unsupervised learning methods, PCA and autoencoder, were also employed. When the output is constrained to two dimensions, both methods can roughly separate configurations of different Wolfram rules into distinct clusters. Once the output is restricted to one dimension, both methods yield an output that shows strong agreement with the density. Due to its simpler learning mechanism, PCA achieves shorter computation time and greater operational efficiency compared to the autoencoder

\begin{acknowledgments} 

 This work is supported in part by the National Key Research and Development Program of China under Grant No.~2024YFA1611003. The National Natural Science Foundation of China (Grant No.~11505071 and 61873104), and the 111 Project 2.0, with Grant No.~BP0820038.
\end{acknowledgments}
 
\bibliographystyle{apsrev4-2}
\bibliography{WCANCML}%

\begin{thebibliography}{41}%
\makeatletter
\providecommand \@ifxundefined [1]{%
 \@ifx{#1\undefined}
}%
\providecommand \@ifnum [1]{%
 \ifnum #1\expandafter \@firstoftwo
 \else \expandafter \@secondoftwo
 \fi
}%
\providecommand \@ifx [1]{%
 \ifx #1\expandafter \@firstoftwo
 \else \expandafter \@secondoftwo
 \fi
}%
\providecommand \natexlab [1]{#1}%
\providecommand \enquote  [1]{``#1''}%
\providecommand \bibnamefont  [1]{#1}%
\providecommand \bibfnamefont [1]{#1}%
\providecommand \citenamefont [1]{#1}%
\providecommand \href@noop [0]{\@secondoftwo}%
\providecommand \href [0]{\begingroup \@sanitize@url \@href}%
\providecommand \@href[1]{\@@startlink{#1}\@@href}%
\providecommand \@@href[1]{\endgroup#1\@@endlink}%
\providecommand \@sanitize@url [0]{\catcode `\\12\catcode `\$12\catcode `\&12\catcode `\#12\catcode `\^12\catcode `\_12\catcode `\%12\relax}%
\providecommand \@@startlink[1]{}%
\providecommand \@@endlink[0]{}%
\providecommand \url  [0]{\begingroup\@sanitize@url \@url }%
\providecommand \@url [1]{\endgroup\@href {#1}{\urlprefix }}%
\providecommand \urlprefix  [0]{URL }%
\providecommand \Eprint [0]{\href }%
\providecommand \doibase [0]{https://doi.org/}%
\providecommand \selectlanguage [0]{\@gobble}%
\providecommand \bibinfo  [0]{\@secondoftwo}%
\providecommand \bibfield  [0]{\@secondoftwo}%
\providecommand \translation [1]{[#1]}%
\providecommand \BibitemOpen [0]{}%
\providecommand \bibitemStop [0]{}%
\providecommand \bibitemNoStop [0]{.\EOS\space}%
\providecommand \EOS [0]{\spacefactor3000\relax}%
\providecommand \BibitemShut  [1]{\csname bibitem#1\endcsname}%
\let\auto@bib@innerbib\@empty
\bibitem [{\citenamefont {Conway}\ \emph {et~al.}(1970)\citenamefont {Conway} \emph {et~al.}}]{conway1970game}%
  \BibitemOpen
  \bibfield  {author} {\bibinfo {author} {\bibfnamefont {J.}~\bibnamefont {Conway}} \emph {et~al.},\ }\href@noop {} {\bibfield  {journal} {\bibinfo  {journal} {Scientific American}\ }\textbf {\bibinfo {volume} {223}},\ \bibinfo {pages} {4} (\bibinfo {year} {1970})}\BibitemShut {NoStop}%
\bibitem [{\citenamefont {Wolfram}(1983)}]{wolfram1983statistical}%
  \BibitemOpen
  \bibfield  {author} {\bibinfo {author} {\bibfnamefont {S.}~\bibnamefont {Wolfram}},\ }\href@noop {} {\bibfield  {journal} {\bibinfo  {journal} {Reviews of modern physics}\ }\textbf {\bibinfo {volume} {55}},\ \bibinfo {pages} {601} (\bibinfo {year} {1983})}\BibitemShut {NoStop}%
\bibitem [{\citenamefont {Domany}\ and\ \citenamefont {Kinzel}(1984)}]{domany1984equivalence}%
  \BibitemOpen
  \bibfield  {author} {\bibinfo {author} {\bibfnamefont {E.}~\bibnamefont {Domany}}\ and\ \bibinfo {author} {\bibfnamefont {W.}~\bibnamefont {Kinzel}},\ }\href@noop {} {\bibfield  {journal} {\bibinfo  {journal} {Physical review letters}\ }\textbf {\bibinfo {volume} {53}},\ \bibinfo {pages} {311} (\bibinfo {year} {1984})}\BibitemShut {NoStop}%
\bibitem [{\citenamefont {Ilachinski}(2001)}]{ilachinski2001cellular}%
  \BibitemOpen
  \bibfield  {author} {\bibinfo {author} {\bibfnamefont {A.}~\bibnamefont {Ilachinski}},\ }\href@noop {} {\emph {\bibinfo {title} {Cellular automata: a discrete universe}}}\ (\bibinfo  {publisher} {World Scientific Publishing Company},\ \bibinfo {year} {2001})\BibitemShut {NoStop}%
\bibitem [{\citenamefont {Grimm}\ and\ \citenamefont {Railsback}(2005)}]{Grimm2005}%
  \BibitemOpen
  \bibfield  {author} {\bibinfo {author} {\bibfnamefont {V.}~\bibnamefont {Grimm}}\ and\ \bibinfo {author} {\bibfnamefont {S.~F.}\ \bibnamefont {Railsback}},\ }\href@noop {} {\bibfield  {journal} {\bibinfo  {journal} {Princeton University Press}\ } (\bibinfo {year} {2005})}\BibitemShut {NoStop}%
\bibitem [{\citenamefont {Rendell}(2002)}]{rendell2002turing}%
  \BibitemOpen
  \bibfield  {author} {\bibinfo {author} {\bibfnamefont {P.}~\bibnamefont {Rendell}},\ }in\ \href@noop {} {\emph {\bibinfo {booktitle} {Collision-based computing}}}\ (\bibinfo  {publisher} {Springer},\ \bibinfo {year} {2002})\ pp.\ \bibinfo {pages} {513--539}\BibitemShut {NoStop}%
\bibitem [{\citenamefont {Batty}(1997)}]{batty1997cellular}%
  \BibitemOpen
  \bibfield  {author} {\bibinfo {author} {\bibfnamefont {M.}~\bibnamefont {Batty}},\ }\href@noop {} {\bibfield  {journal} {\bibinfo  {journal} {Journal of the American planning association}\ }\textbf {\bibinfo {volume} {63}},\ \bibinfo {pages} {266} (\bibinfo {year} {1997})}\BibitemShut {NoStop}%
\bibitem [{\citenamefont {Nagel}\ and\ \citenamefont {Schreckenberg}(1992)}]{nagel1992cellular}%
  \BibitemOpen
  \bibfield  {author} {\bibinfo {author} {\bibfnamefont {K.}~\bibnamefont {Nagel}}\ and\ \bibinfo {author} {\bibfnamefont {M.}~\bibnamefont {Schreckenberg}},\ }\href@noop {} {\bibfield  {journal} {\bibinfo  {journal} {Journal de physique I}\ }\textbf {\bibinfo {volume} {2}},\ \bibinfo {pages} {2221} (\bibinfo {year} {1992})}\BibitemShut {NoStop}%
\bibitem [{\citenamefont {Yacoubi}(2008)}]{yacoubi2008mathematical}%
  \BibitemOpen
  \bibfield  {author} {\bibinfo {author} {\bibfnamefont {S.~E.}\ \bibnamefont {Yacoubi}},\ }\href@noop {} {\bibfield  {journal} {\bibinfo  {journal} {International journal of systems science}\ }\textbf {\bibinfo {volume} {39}},\ \bibinfo {pages} {529} (\bibinfo {year} {2008})}\BibitemShut {NoStop}%
\bibitem [{\citenamefont {Bartolozzi}\ and\ \citenamefont {Thomas}(2004)}]{bartolozzi2004stochastic}%
  \BibitemOpen
  \bibfield  {author} {\bibinfo {author} {\bibfnamefont {M.}~\bibnamefont {Bartolozzi}}\ and\ \bibinfo {author} {\bibfnamefont {A.~W.}\ \bibnamefont {Thomas}},\ }\href@noop {} {\bibfield  {journal} {\bibinfo  {journal} {Physical Review E—Statistical, Nonlinear, and Soft Matter Physics}\ }\textbf {\bibinfo {volume} {69}},\ \bibinfo {pages} {046112} (\bibinfo {year} {2004})}\BibitemShut {NoStop}%
\bibitem [{\citenamefont {Zhao}\ \emph {et~al.}(2009)\citenamefont {Zhao}, \citenamefont {Billings},\ and\ \citenamefont {Coca}}]{zhao2009cellular}%
  \BibitemOpen
  \bibfield  {author} {\bibinfo {author} {\bibfnamefont {Y.}~\bibnamefont {Zhao}}, \bibinfo {author} {\bibfnamefont {S.~A.}\ \bibnamefont {Billings}},\ and\ \bibinfo {author} {\bibfnamefont {D.}~\bibnamefont {Coca}},\ }\href@noop {} {\bibfield  {journal} {\bibinfo  {journal} {International Journal of Modelling, Identification and Control}\ }\textbf {\bibinfo {volume} {6}},\ \bibinfo {pages} {119} (\bibinfo {year} {2009})}\BibitemShut {NoStop}%
\bibitem [{\citenamefont {Ermentrout}\ and\ \citenamefont {Edelstein-Keshet}(1993)}]{ermentrout1993cellular}%
  \BibitemOpen
  \bibfield  {author} {\bibinfo {author} {\bibfnamefont {G.~B.}\ \bibnamefont {Ermentrout}}\ and\ \bibinfo {author} {\bibfnamefont {L.}~\bibnamefont {Edelstein-Keshet}},\ }\href@noop {} {\bibfield  {journal} {\bibinfo  {journal} {Journal of theoretical Biology}\ }\textbf {\bibinfo {volume} {160}},\ \bibinfo {pages} {97} (\bibinfo {year} {1993})}\BibitemShut {NoStop}%
\bibitem [{\citenamefont {Lai}(2019)}]{lai2019comparison}%
  \BibitemOpen
  \bibfield  {author} {\bibinfo {author} {\bibfnamefont {Y.}~\bibnamefont {Lai}},\ }in\ \href@noop {} {\emph {\bibinfo {booktitle} {Journal of Physics: Conference Series}}},\ Vol.\ \bibinfo {volume} {1314}\ (\bibinfo {organization} {IOP Publishing},\ \bibinfo {year} {2019})\ p.\ \bibinfo {pages} {012148}\BibitemShut {NoStop}%
\bibitem [{\citenamefont {Pathak}\ \emph {et~al.}(2018)\citenamefont {Pathak}, \citenamefont {Pandey},\ and\ \citenamefont {Rautaray}}]{pathak2018application}%
  \BibitemOpen
  \bibfield  {author} {\bibinfo {author} {\bibfnamefont {A.~R.}\ \bibnamefont {Pathak}}, \bibinfo {author} {\bibfnamefont {M.}~\bibnamefont {Pandey}},\ and\ \bibinfo {author} {\bibfnamefont {S.}~\bibnamefont {Rautaray}},\ }\href@noop {} {\bibfield  {journal} {\bibinfo  {journal} {Procedia computer science}\ }\textbf {\bibinfo {volume} {132}},\ \bibinfo {pages} {1706} (\bibinfo {year} {2018})}\BibitemShut {NoStop}%
\bibitem [{\citenamefont {Graves}\ \emph {et~al.}(2013)\citenamefont {Graves}, \citenamefont {Mohamed},\ and\ \citenamefont {Hinton}}]{Hinton}%
  \BibitemOpen
  \bibfield  {author} {\bibinfo {author} {\bibfnamefont {A.}~\bibnamefont {Graves}}, \bibinfo {author} {\bibfnamefont {A.-r.}\ \bibnamefont {Mohamed}},\ and\ \bibinfo {author} {\bibfnamefont {G.}~\bibnamefont {Hinton}},\ }in\ \href@noop {} {\emph {\bibinfo {booktitle} {2013 IEEE international conference on acoustics, speech and signal processing}}}\ (\bibinfo {organization} {Ieee},\ \bibinfo {year} {2013})\ pp.\ \bibinfo {pages} {6645--6649}\BibitemShut {NoStop}%
\bibitem [{\citenamefont {Da’u}\ and\ \citenamefont {Salim}(2020)}]{da2020recommendation}%
  \BibitemOpen
  \bibfield  {author} {\bibinfo {author} {\bibfnamefont {A.}~\bibnamefont {Da’u}}\ and\ \bibinfo {author} {\bibfnamefont {N.}~\bibnamefont {Salim}},\ }\href@noop {} {\bibfield  {journal} {\bibinfo  {journal} {Artificial Intelligence Review}\ }\textbf {\bibinfo {volume} {53}},\ \bibinfo {pages} {2709} (\bibinfo {year} {2020})}\BibitemShut {NoStop}%
\bibitem [{\citenamefont {Hansen}(2020)}]{hansen2020virtue}%
  \BibitemOpen
  \bibfield  {author} {\bibinfo {author} {\bibfnamefont {K.~B.}\ \bibnamefont {Hansen}},\ }\href@noop {} {\bibfield  {journal} {\bibinfo  {journal} {Big Data \& Society}\ }\textbf {\bibinfo {volume} {7}},\ \bibinfo {pages} {2053951720926558} (\bibinfo {year} {2020})}\BibitemShut {NoStop}%
\bibitem [{\citenamefont {Catacutan}\ \emph {et~al.}(2024)\citenamefont {Catacutan}, \citenamefont {Alexander}, \citenamefont {Arnold},\ and\ \citenamefont {Stokes}}]{catacutan2024machine}%
  \BibitemOpen
  \bibfield  {author} {\bibinfo {author} {\bibfnamefont {D.~B.}\ \bibnamefont {Catacutan}}, \bibinfo {author} {\bibfnamefont {J.}~\bibnamefont {Alexander}}, \bibinfo {author} {\bibfnamefont {A.}~\bibnamefont {Arnold}},\ and\ \bibinfo {author} {\bibfnamefont {J.~M.}\ \bibnamefont {Stokes}},\ }\href@noop {} {\bibfield  {journal} {\bibinfo  {journal} {Nature Chemical Biology}\ }\textbf {\bibinfo {volume} {20}},\ \bibinfo {pages} {960} (\bibinfo {year} {2024})}\BibitemShut {NoStop}%
\bibitem [{\citenamefont {Whalen}\ \emph {et~al.}(2022)\citenamefont {Whalen}, \citenamefont {Schreiber}, \citenamefont {Noble},\ and\ \citenamefont {Pollard}}]{whalen2022navigating}%
  \BibitemOpen
  \bibfield  {author} {\bibinfo {author} {\bibfnamefont {S.}~\bibnamefont {Whalen}}, \bibinfo {author} {\bibfnamefont {J.}~\bibnamefont {Schreiber}}, \bibinfo {author} {\bibfnamefont {W.~S.}\ \bibnamefont {Noble}},\ and\ \bibinfo {author} {\bibfnamefont {K.~S.}\ \bibnamefont {Pollard}},\ }\href@noop {} {\bibfield  {journal} {\bibinfo  {journal} {Nature Reviews Genetics}\ }\textbf {\bibinfo {volume} {23}},\ \bibinfo {pages} {169} (\bibinfo {year} {2022})}\BibitemShut {NoStop}%
\bibitem [{\citenamefont {Du}\ \emph {et~al.}(2023)\citenamefont {Du}, \citenamefont {Watkins}, \citenamefont {Wang}, \citenamefont {Colas}, \citenamefont {Darrell}, \citenamefont {Abbeel}, \citenamefont {Gupta},\ and\ \citenamefont {Andreas}}]{du2023guiding}%
  \BibitemOpen
  \bibfield  {author} {\bibinfo {author} {\bibfnamefont {Y.}~\bibnamefont {Du}}, \bibinfo {author} {\bibfnamefont {O.}~\bibnamefont {Watkins}}, \bibinfo {author} {\bibfnamefont {Z.}~\bibnamefont {Wang}}, \bibinfo {author} {\bibfnamefont {C.}~\bibnamefont {Colas}}, \bibinfo {author} {\bibfnamefont {T.}~\bibnamefont {Darrell}}, \bibinfo {author} {\bibfnamefont {P.}~\bibnamefont {Abbeel}}, \bibinfo {author} {\bibfnamefont {A.}~\bibnamefont {Gupta}},\ and\ \bibinfo {author} {\bibfnamefont {J.}~\bibnamefont {Andreas}},\ }in\ \href@noop {} {\emph {\bibinfo {booktitle} {International Conference on Machine Learning}}}\ (\bibinfo {organization} {PMLR},\ \bibinfo {year} {2023})\ pp.\ \bibinfo {pages} {8657--8677}\BibitemShut {NoStop}%
\bibitem [{\citenamefont {Tani}\ \emph {et~al.}(2021)\citenamefont {Tani}, \citenamefont {Rand}, \citenamefont {Veelken},\ and\ \citenamefont {Kadastik}}]{tani2021evolutionary}%
  \BibitemOpen
  \bibfield  {author} {\bibinfo {author} {\bibfnamefont {L.}~\bibnamefont {Tani}}, \bibinfo {author} {\bibfnamefont {D.}~\bibnamefont {Rand}}, \bibinfo {author} {\bibfnamefont {C.}~\bibnamefont {Veelken}},\ and\ \bibinfo {author} {\bibfnamefont {M.}~\bibnamefont {Kadastik}},\ }\href@noop {} {\bibfield  {journal} {\bibinfo  {journal} {The European Physical Journal C}\ }\textbf {\bibinfo {volume} {81}},\ \bibinfo {pages} {170} (\bibinfo {year} {2021})}\BibitemShut {NoStop}%
\bibitem [{\citenamefont {Ma}\ \emph {et~al.}(2023)\citenamefont {Ma}, \citenamefont {Liu},\ and\ \citenamefont {Li}}]{ma2023jet}%
  \BibitemOpen
  \bibfield  {author} {\bibinfo {author} {\bibfnamefont {F.}~\bibnamefont {Ma}}, \bibinfo {author} {\bibfnamefont {F.}~\bibnamefont {Liu}},\ and\ \bibinfo {author} {\bibfnamefont {W.}~\bibnamefont {Li}},\ }\href@noop {} {\bibfield  {journal} {\bibinfo  {journal} {Physical Review D}\ }\textbf {\bibinfo {volume} {108}},\ \bibinfo {pages} {072007} (\bibinfo {year} {2023})}\BibitemShut {NoStop}%
\bibitem [{\citenamefont {Rodr{\'\i}guez}\ \emph {et~al.}(2022)\citenamefont {Rodr{\'\i}guez}, \citenamefont {Rodr{\'\i}guez-Rodr{\'\i}guez},\ and\ \citenamefont {Woo}}]{rodriguez2022application}%
  \BibitemOpen
  \bibfield  {author} {\bibinfo {author} {\bibfnamefont {J.-V.}\ \bibnamefont {Rodr{\'\i}guez}}, \bibinfo {author} {\bibfnamefont {I.}~\bibnamefont {Rodr{\'\i}guez-Rodr{\'\i}guez}},\ and\ \bibinfo {author} {\bibfnamefont {W.~L.}\ \bibnamefont {Woo}},\ }\href@noop {} {\bibfield  {journal} {\bibinfo  {journal} {Wiley Interdisciplinary Reviews: Data Mining and Knowledge Discovery}\ }\textbf {\bibinfo {volume} {12}},\ \bibinfo {pages} {e1476} (\bibinfo {year} {2022})}\BibitemShut {NoStop}%
\bibitem [{\citenamefont {Qiu}\ \emph {et~al.}(2024)\citenamefont {Qiu}, \citenamefont {Napolitano}, \citenamefont {Borgani}, \citenamefont {Zhong}, \citenamefont {Li}, \citenamefont {Radovich}, \citenamefont {Lin}, \citenamefont {Dolag}, \citenamefont {Tortora}, \citenamefont {Wang} \emph {et~al.}}]{qiu2024cosmology}%
  \BibitemOpen
  \bibfield  {author} {\bibinfo {author} {\bibfnamefont {L.}~\bibnamefont {Qiu}}, \bibinfo {author} {\bibfnamefont {N.~R.}\ \bibnamefont {Napolitano}}, \bibinfo {author} {\bibfnamefont {S.}~\bibnamefont {Borgani}}, \bibinfo {author} {\bibfnamefont {F.}~\bibnamefont {Zhong}}, \bibinfo {author} {\bibfnamefont {X.}~\bibnamefont {Li}}, \bibinfo {author} {\bibfnamefont {M.}~\bibnamefont {Radovich}}, \bibinfo {author} {\bibfnamefont {W.}~\bibnamefont {Lin}}, \bibinfo {author} {\bibfnamefont {K.}~\bibnamefont {Dolag}}, \bibinfo {author} {\bibfnamefont {C.}~\bibnamefont {Tortora}}, \bibinfo {author} {\bibfnamefont {Y.}~\bibnamefont {Wang}}, \emph {et~al.},\ }\href@noop {} {\bibfield  {journal} {\bibinfo  {journal} {Astronomy \& Astrophysics}\ }\textbf {\bibinfo {volume} {687}},\ \bibinfo {pages} {A1} (\bibinfo {year} {2024})}\BibitemShut {NoStop}%
\bibitem [{\citenamefont {Xiao}\ \emph {et~al.}(2022)\citenamefont {Xiao}, \citenamefont {Huang}, \citenamefont {Li}, \citenamefont {Fan},\ and\ \citenamefont {Zeng}}]{xiao2022intelligent}%
  \BibitemOpen
  \bibfield  {author} {\bibinfo {author} {\bibfnamefont {T.}~\bibnamefont {Xiao}}, \bibinfo {author} {\bibfnamefont {J.}~\bibnamefont {Huang}}, \bibinfo {author} {\bibfnamefont {H.}~\bibnamefont {Li}}, \bibinfo {author} {\bibfnamefont {J.}~\bibnamefont {Fan}},\ and\ \bibinfo {author} {\bibfnamefont {G.}~\bibnamefont {Zeng}},\ }\href@noop {} {\bibfield  {journal} {\bibinfo  {journal} {npj Quantum Information}\ }\textbf {\bibinfo {volume} {8}},\ \bibinfo {pages} {138} (\bibinfo {year} {2022})}\BibitemShut {NoStop}%
\bibitem [{\citenamefont {Gibbs}\ \emph {et~al.}(2024)\citenamefont {Gibbs}, \citenamefont {Holmes}, \citenamefont {Caro}, \citenamefont {Ezzell}, \citenamefont {Huang}, \citenamefont {Cincio}, \citenamefont {Sornborger},\ and\ \citenamefont {Coles}}]{gibbs2024dynamical}%
  \BibitemOpen
  \bibfield  {author} {\bibinfo {author} {\bibfnamefont {J.}~\bibnamefont {Gibbs}}, \bibinfo {author} {\bibfnamefont {Z.}~\bibnamefont {Holmes}}, \bibinfo {author} {\bibfnamefont {M.~C.}\ \bibnamefont {Caro}}, \bibinfo {author} {\bibfnamefont {N.}~\bibnamefont {Ezzell}}, \bibinfo {author} {\bibfnamefont {H.-Y.}\ \bibnamefont {Huang}}, \bibinfo {author} {\bibfnamefont {L.}~\bibnamefont {Cincio}}, \bibinfo {author} {\bibfnamefont {A.~T.}\ \bibnamefont {Sornborger}},\ and\ \bibinfo {author} {\bibfnamefont {P.~J.}\ \bibnamefont {Coles}},\ }\href@noop {} {\bibfield  {journal} {\bibinfo  {journal} {Physical Review Research}\ }\textbf {\bibinfo {volume} {6}},\ \bibinfo {pages} {013241} (\bibinfo {year} {2024})}\BibitemShut {NoStop}%
\bibitem [{\citenamefont {Chittoor}\ and\ \citenamefont {Simeone}(2023)}]{chittoor2023quantum}%
  \BibitemOpen
  \bibfield  {author} {\bibinfo {author} {\bibfnamefont {H.~H.~S.}\ \bibnamefont {Chittoor}}\ and\ \bibinfo {author} {\bibfnamefont {O.}~\bibnamefont {Simeone}},\ }\href@noop {} {\bibfield  {journal} {\bibinfo  {journal} {Entropy}\ }\textbf {\bibinfo {volume} {25}},\ \bibinfo {pages} {352} (\bibinfo {year} {2023})}\BibitemShut {NoStop}%
\bibitem [{\citenamefont {Carrasquilla}\ and\ \citenamefont {Melko}(2017)}]{Carrasquilla}%
  \BibitemOpen
  \bibfield  {author} {\bibinfo {author} {\bibfnamefont {J.}~\bibnamefont {Carrasquilla}}\ and\ \bibinfo {author} {\bibfnamefont {R.~G.}\ \bibnamefont {Melko}},\ }\href@noop {} {\bibfield  {journal} {\bibinfo  {journal} {Nature Physics}\ }\textbf {\bibinfo {volume} {13}},\ \bibinfo {pages} {431} (\bibinfo {year} {2017})}\BibitemShut {NoStop}%
\bibitem [{\citenamefont {Zhang}\ \emph {et~al.}(2019)\citenamefont {Zhang}, \citenamefont {Wei}, \citenamefont {Zhang}, \citenamefont {Zhu},\ and\ \citenamefont {Chang}}]{Zhang3DIsing}%
  \BibitemOpen
  \bibfield  {author} {\bibinfo {author} {\bibfnamefont {R.}~\bibnamefont {Zhang}}, \bibinfo {author} {\bibfnamefont {B.}~\bibnamefont {Wei}}, \bibinfo {author} {\bibfnamefont {D.}~\bibnamefont {Zhang}}, \bibinfo {author} {\bibfnamefont {J.-J.}\ \bibnamefont {Zhu}},\ and\ \bibinfo {author} {\bibfnamefont {K.}~\bibnamefont {Chang}},\ }\href@noop {} {\bibfield  {journal} {\bibinfo  {journal} {Physical Review B}\ }\textbf {\bibinfo {volume} {99}},\ \bibinfo {pages} {094427} (\bibinfo {year} {2019})}\BibitemShut {NoStop}%
\bibitem [{\citenamefont {Hu}\ \emph {et~al.}(2023)\citenamefont {Hu}, \citenamefont {Sun}, \citenamefont {Liu}, \citenamefont {Zhang}, \citenamefont {Liu}, \citenamefont {Fan}, \citenamefont {Chen},\ and\ \citenamefont {Chen}}]{hu2023universality}%
  \BibitemOpen
  \bibfield  {author} {\bibinfo {author} {\bibfnamefont {G.}~\bibnamefont {Hu}}, \bibinfo {author} {\bibfnamefont {Y.}~\bibnamefont {Sun}}, \bibinfo {author} {\bibfnamefont {T.}~\bibnamefont {Liu}}, \bibinfo {author} {\bibfnamefont {Y.}~\bibnamefont {Zhang}}, \bibinfo {author} {\bibfnamefont {M.}~\bibnamefont {Liu}}, \bibinfo {author} {\bibfnamefont {J.}~\bibnamefont {Fan}}, \bibinfo {author} {\bibfnamefont {W.}~\bibnamefont {Chen}},\ and\ \bibinfo {author} {\bibfnamefont {X.}~\bibnamefont {Chen}},\ }\href@noop {} {\bibfield  {journal} {\bibinfo  {journal} {Science China Physics, Mechanics \& Astronomy}\ }\textbf {\bibinfo {volume} {66}},\ \bibinfo {pages} {120511} (\bibinfo {year} {2023})}\BibitemShut {NoStop}%
\bibitem [{\citenamefont {K{\"a}ming}\ \emph {et~al.}(2021)\citenamefont {K{\"a}ming}, \citenamefont {Dawid}, \citenamefont {Kottmann}, \citenamefont {Lewenstein}, \citenamefont {Sengstock}, \citenamefont {Dauphin},\ and\ \citenamefont {Weitenberg}}]{kaming2021unsupervised}%
  \BibitemOpen
  \bibfield  {author} {\bibinfo {author} {\bibfnamefont {N.}~\bibnamefont {K{\"a}ming}}, \bibinfo {author} {\bibfnamefont {A.}~\bibnamefont {Dawid}}, \bibinfo {author} {\bibfnamefont {K.}~\bibnamefont {Kottmann}}, \bibinfo {author} {\bibfnamefont {M.}~\bibnamefont {Lewenstein}}, \bibinfo {author} {\bibfnamefont {K.}~\bibnamefont {Sengstock}}, \bibinfo {author} {\bibfnamefont {A.}~\bibnamefont {Dauphin}},\ and\ \bibinfo {author} {\bibfnamefont {C.}~\bibnamefont {Weitenberg}},\ }\href@noop {} {\bibfield  {journal} {\bibinfo  {journal} {Machine Learning: Science and Technology}\ }\textbf {\bibinfo {volume} {2}},\ \bibinfo {pages} {035037} (\bibinfo {year} {2021})}\BibitemShut {NoStop}%
\bibitem [{\citenamefont {Holanda}\ and\ \citenamefont {Griffith}(2020)}]{holanda2020machine}%
  \BibitemOpen
  \bibfield  {author} {\bibinfo {author} {\bibfnamefont {N.}~\bibnamefont {Holanda}}\ and\ \bibinfo {author} {\bibfnamefont {M.}~\bibnamefont {Griffith}},\ }\href@noop {} {\bibfield  {journal} {\bibinfo  {journal} {Physical Review B}\ }\textbf {\bibinfo {volume} {102}},\ \bibinfo {pages} {054107} (\bibinfo {year} {2020})}\BibitemShut {NoStop}%
\bibitem [{\citenamefont {Tuo}\ \emph {et~al.}(2024)\citenamefont {Tuo}, \citenamefont {Li}, \citenamefont {Deng},\ and\ \citenamefont {Zhu}}]{tuo2024supervised}%
  \BibitemOpen
  \bibfield  {author} {\bibinfo {author} {\bibfnamefont {K.}~\bibnamefont {Tuo}}, \bibinfo {author} {\bibfnamefont {W.}~\bibnamefont {Li}}, \bibinfo {author} {\bibfnamefont {S.}~\bibnamefont {Deng}},\ and\ \bibinfo {author} {\bibfnamefont {Y.}~\bibnamefont {Zhu}},\ }\href@noop {} {\bibfield  {journal} {\bibinfo  {journal} {Physical Review E}\ }\textbf {\bibinfo {volume} {110}},\ \bibinfo {pages} {024102} (\bibinfo {year} {2024})}\BibitemShut {NoStop}%
\bibitem [{\citenamefont {Wang}\ \emph {et~al.}(2024)\citenamefont {Wang}, \citenamefont {Li}, \citenamefont {Liu},\ and\ \citenamefont {Shen}}]{wang2024supervised}%
  \BibitemOpen
  \bibfield  {author} {\bibinfo {author} {\bibfnamefont {Y.}~\bibnamefont {Wang}}, \bibinfo {author} {\bibfnamefont {W.}~\bibnamefont {Li}}, \bibinfo {author} {\bibfnamefont {F.}~\bibnamefont {Liu}},\ and\ \bibinfo {author} {\bibfnamefont {J.}~\bibnamefont {Shen}},\ }\href@noop {} {\bibfield  {journal} {\bibinfo  {journal} {Machine Learning: Science and Technology}\ }\textbf {\bibinfo {volume} {5}},\ \bibinfo {pages} {015033} (\bibinfo {year} {2024})}\BibitemShut {NoStop}%
\bibitem [{\citenamefont {Grassberger}(1983)}]{grassberger1983new}%
  \BibitemOpen
  \bibfield  {author} {\bibinfo {author} {\bibfnamefont {P.}~\bibnamefont {Grassberger}},\ }\href@noop {} {\bibfield  {journal} {\bibinfo  {journal} {Physical Review A}\ }\textbf {\bibinfo {volume} {28}},\ \bibinfo {pages} {3666} (\bibinfo {year} {1983})}\BibitemShut {NoStop}%
\bibitem [{\citenamefont {Liou}\ \emph {et~al.}(2014)\citenamefont {Liou}, \citenamefont {Cheng}, \citenamefont {Liou},\ and\ \citenamefont {Liou}}]{liou2014autoencoder}%
  \BibitemOpen
  \bibfield  {author} {\bibinfo {author} {\bibfnamefont {C.-Y.}\ \bibnamefont {Liou}}, \bibinfo {author} {\bibfnamefont {W.-C.}\ \bibnamefont {Cheng}}, \bibinfo {author} {\bibfnamefont {J.-W.}\ \bibnamefont {Liou}},\ and\ \bibinfo {author} {\bibfnamefont {D.-R.}\ \bibnamefont {Liou}},\ }\href@noop {} {\bibfield  {journal} {\bibinfo  {journal} {Neurocomputing}\ }\textbf {\bibinfo {volume} {139}},\ \bibinfo {pages} {84} (\bibinfo {year} {2014})}\BibitemShut {NoStop}%
\bibitem [{\citenamefont {Pinheiro~Cinelli}\ \emph {et~al.}(2021)\citenamefont {Pinheiro~Cinelli}, \citenamefont {Ara{\'u}jo~Marins}, \citenamefont {Barros~da Silva},\ and\ \citenamefont {Lima~Netto}}]{pinheiro2021variational}%
  \BibitemOpen
  \bibfield  {author} {\bibinfo {author} {\bibfnamefont {L.}~\bibnamefont {Pinheiro~Cinelli}}, \bibinfo {author} {\bibfnamefont {M.}~\bibnamefont {Ara{\'u}jo~Marins}}, \bibinfo {author} {\bibfnamefont {E.~A.}\ \bibnamefont {Barros~da Silva}},\ and\ \bibinfo {author} {\bibfnamefont {S.}~\bibnamefont {Lima~Netto}},\ }in\ \href@noop {} {\emph {\bibinfo {booktitle} {Variational methods for machine learning with applications to deep networks}}}\ (\bibinfo  {publisher} {Springer},\ \bibinfo {year} {2021})\ pp.\ \bibinfo {pages} {111--149}\BibitemShut {NoStop}%
\bibitem [{\citenamefont {Pinaya}\ \emph {et~al.}(2020)\citenamefont {Pinaya}, \citenamefont {Vieira}, \citenamefont {Garcia-Dias},\ and\ \citenamefont {Mechelli}}]{pinaya2020autoencoders}%
  \BibitemOpen
  \bibfield  {author} {\bibinfo {author} {\bibfnamefont {W.~H.~L.}\ \bibnamefont {Pinaya}}, \bibinfo {author} {\bibfnamefont {S.}~\bibnamefont {Vieira}}, \bibinfo {author} {\bibfnamefont {R.}~\bibnamefont {Garcia-Dias}},\ and\ \bibinfo {author} {\bibfnamefont {A.}~\bibnamefont {Mechelli}},\ }in\ \href@noop {} {\emph {\bibinfo {booktitle} {Machine learning}}}\ (\bibinfo  {publisher} {Elsevier},\ \bibinfo {year} {2020})\ pp.\ \bibinfo {pages} {193--208}\BibitemShut {NoStop}%
\bibitem [{\citenamefont {Ma{\'c}kiewicz}\ and\ \citenamefont {Ratajczak}(1993)}]{mackiewicz1993principal}%
  \BibitemOpen
  \bibfield  {author} {\bibinfo {author} {\bibfnamefont {A.}~\bibnamefont {Ma{\'c}kiewicz}}\ and\ \bibinfo {author} {\bibfnamefont {W.}~\bibnamefont {Ratajczak}},\ }\href@noop {} {\bibfield  {journal} {\bibinfo  {journal} {Computers \& Geosciences}\ }\textbf {\bibinfo {volume} {19}},\ \bibinfo {pages} {303} (\bibinfo {year} {1993})}\BibitemShut {NoStop}%
\bibitem [{\citenamefont {Greenacre}\ \emph {et~al.}(2022)\citenamefont {Greenacre}, \citenamefont {Groenen}, \citenamefont {Hastie}, \citenamefont {d’Enza}, \citenamefont {Markos},\ and\ \citenamefont {Tuzhilina}}]{greenacre2022principal}%
  \BibitemOpen
  \bibfield  {author} {\bibinfo {author} {\bibfnamefont {M.}~\bibnamefont {Greenacre}}, \bibinfo {author} {\bibfnamefont {P.~J.}\ \bibnamefont {Groenen}}, \bibinfo {author} {\bibfnamefont {T.}~\bibnamefont {Hastie}}, \bibinfo {author} {\bibfnamefont {A.~I.}\ \bibnamefont {d’Enza}}, \bibinfo {author} {\bibfnamefont {A.}~\bibnamefont {Markos}},\ and\ \bibinfo {author} {\bibfnamefont {E.}~\bibnamefont {Tuzhilina}},\ }\href@noop {} {\bibfield  {journal} {\bibinfo  {journal} {Nature Reviews Methods Primers}\ }\textbf {\bibinfo {volume} {2}},\ \bibinfo {pages} {100} (\bibinfo {year} {2022})}\BibitemShut {NoStop}%
\bibitem [{\citenamefont {Bharadiya}(2023)}]{bharadiya2023tutorial}%
  \BibitemOpen
  \bibfield  {author} {\bibinfo {author} {\bibfnamefont {J.~P.}\ \bibnamefont {Bharadiya}},\ }\href@noop {} {\bibfield  {journal} {\bibinfo  {journal} {International journal of innovative science and research technology}\ }\textbf {\bibinfo {volume} {8}},\ \bibinfo {pages} {2028} (\bibinfo {year} {2023})}\BibitemShut {NoStop}%
\end{thebibliography}%
\end{document}